\documentclass[journal]{IEEEtran}
\usepackage{amscd,amsmath,amssymb,amsfonts,latexsym,mathrsfs,amsthm}
\usepackage{graphicx} 
\usepackage{setspace} 
\usepackage{graphics,epsfig}
\usepackage[english]{babel}
\usepackage{amsmath}
\usepackage{amsfonts}
\usepackage{amssymb,amsthm}
\usepackage{bm}
\usepackage{mathrsfs}
\usepackage{subfigure}
\usepackage[usenames]{color}
\usepackage{rotating}
\usepackage{url}

\usepackage{colortbl}

\definecolor{MYCOLOR0}{rgb}{0.92,0.92,0.92}

\newtheorem{theo}{Theorem}

\usepackage{flushend}
\usepackage{verbatim}
\usepackage{tabularx}
\usepackage{multirow} 
\usepackage{arydshln}

\UseRawInputEncoding

\ifCLASSINFOpdf
\else
\fi

\begin{document}
%
\title{Compressed Monte Carlo \\ 
 with application in particle filtering 
}
%
%
%

\author{Luca Martino$^{\top}$, V{\'i}ctor Elvira$^*$ \\
{\small $^{\top}$ Dep. of Signal Processing, Universidad Rey Juan Carlos (URJC) and Universidad Carlos III de Madrid (UC3M)} \\
{\small$^*$ IMT Lille Douai, Cit{\'e} Scientifique, Rue Guglielmo Marconi, 20145, Villeneuve dÕAscq 59653, (France)} \\
\thanks{E-mail: {\tt luca.martino@urjc.es}.}
}

\maketitle

\begin{abstract}
Bayesian models have become very popular over the last years in several fields such as signal processing, statistics, and machine learning.  Bayesian inference requires the approximation of complicated integrals involving posterior distributions.  
For this purpose, Monte Carlo (MC) methods, such as Markov Chain Monte Carlo and importance sampling algorithms, are often employed. In this work, we introduce {the theory} and practice of a Compressed MC (C-MC) scheme {to} compress the  {statistical} information contained in a {set} of {random} samples. In its basic version, C-MC is strictly related to the stratification technique, a well-known method used for variance reduction purposes. Deterministic C-MC schemes are also presented, which provide very good performance.
  The compression problem is strictly related to {the moment} matching approach applied in different filtering {techniques, usually called} as Gaussian quadrature rules or sigma-point methods.   C-MC can be employed in a distributed Bayesian inference  {framework} when cheap and fast communications with a central processor are required. Furthermore, C-MC is useful within particle filtering and adaptive IS algorithms, as shown by three novel schemes introduced in this work.  
Six numerical results confirm the benefits of the introduced schemes, outperforming the corresponding benchmark methods. {A related code} is also provided.\footnote{{The  code} is provided at \url{http://www.lucamartino.altervista.org/CMC_CODE_pub_EX1.zip}}
\end{abstract}


\section{Introduction}


An essential problem in signal processing, statistics, and machine learning is the estimation of unknown parameters in probabilistic models from noisy observations. Within the Bayesian inference framework, these problems are addressed by constructing posterior probability density functions (pdfs) of the unknowns \cite{7974876,Sarkka13bo}. 
Unfortunately, the computation of statistical quantities related to these posterior distributions (such as moments or credible intervals)  is analytically impossible in most real-world applications. As a consequence, {the design of efficient computational} algorithms is of utmost interest. Monte Carlo (MC) techniques come to the rescue for solving {the most} difficult problems of inference \cite{Liu04b,Robert04}. They are  {benchmark} tools for approximating complicated integrals involving sophisticated multidimensional target densities,  based on  { drawing of random samples} \cite{Robert04,LucaJesse1}. Markov Chain Monte Carlo (MCMC) algorithms, Importance Sampling (IS) schemes{, and} its sequential version (particle filtering) are the most important classes of MC methods  \cite{Sarkka13bo}.  
\newline
{\bf Determinism and support points.} In order to reduce the computational demand of the Monte Carlo methods and the variance of the corresponding estimators, deterministic procedures have been included within the sampling algorithms. In the so-called variance reduction techniques (e.g., conditioning, stratification, antithetic sampling, and control variates), { negative correlation is induced among the generated samples}, hence obtaining more efficient estimators \cite{mcbookOwen,Wilson84}. In Quasi-Monte Carlo (QMC) methods, deterministic sequences of samples are employed{, based} on the concept of {\it low-discrepancy}, avoiding all {kinds} of randomness \cite{Fearnhead05,SQMC,Niederreiter92}. 
 In the same line, deterministic approximations of the posterior distribution based on quadrature, cubature rules{, or} unscented transformations are often applied, when are available \cite{CKF09,Julier04,Wu06,Sarkka13bo}. 
 These techniques provide a set of particles deterministically chosen (often called {\it sigma points}), { to} match perfectly the estimation of a pre-established number of moments of the posterior density. Most of them are derived {for integrals that involve a} Gaussian distribution \cite{Sarkka13bo}. 
 These techniques are usually used in filtering applications as {an extension} of the standard Kalman filtering and as {an alternative} to the particle filtering techniques based on MC sampling. The quadrature rules are very efficient since with $N$ weighted particles summarized exactly the first $2N$ non-central moments. However, quadrature approximations are available only for certain target densities. Indeed, the true values of the moments must be known and a solution of {a highly} non-linear system must be provided. {  This} is possible only for specific target densities. 
More generally, the idea of sigma points is strictly connected to the need of {\it summarizing} a given distribution (and/or function) with a set of {\it representative, support points}, deterministically selected \cite{SuppPoints,ProSuppPoints}. This is an important topic is in computational statistics and has gained increasing attention in the last years: some relevant examples are the herding algorithms  \cite{SteinPoints,Chen10,Lacoste15,Huszar_2012}, the studies about the representative points previously mentioned \cite{ProSuppPoints,SuppPoints}, as well as 
{space-filling} and experimental designs \cite{Pronzato17}. Some of them have been applied jointly with MC schemes or used for numerical integration problems \cite{Lacoste15,Huszar_2012}. 
 \newline 
  \newline
{\bf Contribution.} In this work, we introduce different schemes for compressing the information contained in $N$ Monte Carlo samples into $M<N$ weighted particles. { They are} based on the {so-called} stratification approach \cite{mcbookOwen,Robert04}. In the Compressed Monte Carlo (C-MC) schemes, we replace the particle MC approximation obtained by $N$ unweighted samples (e.g., generated by an MCMC algorithm) or weighted samples (e.g., generated by an IS algorithm), with another particle approximation with $M<N$ summary weighted samples.  {We} desire to reduce the loss of information in terms of moment matching, in the same fashion of the quadrature rules. In this sense, the $M$ summary particles can be considered as approximate sigma points. Furthermore, for a specific choice of the partition ({specifically,} see the case of unweighted C-MC samples in Section \ref{ChoicePartition}), an approximate low-discrepancy sequence is obtained{, i.e., a QMC sequence is generated}. 
Several alternatives and extensions are presented, including the random or deterministic selection of the summary particles.  
\newline 
 The  C-MC approach has a direct application in a parallel or distributed Bayesian framework {with a centralized node}, as discussed in Section \ref{DistrSect2} and graphically represented in Figure \ref{Fig1teo} {. In this scenario,} different local low-power nodes must transmit to a central node the results of their local Bayesian analysis, { to} provide a common complete inference  \cite{EmbaraMCMC,Bolic05,Read2014}. The transmission should have the minimum possible cost and contain the maximum amount of information. 
 Hence, the information must be properly compressed before {being} transmitted (see Section \ref{DistrSect} for further details).
 C-MC {can be considered} an improvement of the bootstrap strategy, applied in different works regarding parallel sequential Monte Carlo schemes, where several resampled particles are transmitted jointly with {a} proper aggregated weight \cite{Bolic05,Read2014,Pisland15,GIS18}.  { However, the range of application of C-MC is not only restricted to the distributed scenario. We introduce two novel particle filtering schemes based on the C-MC approach. The first scheme enhances the well-known Gaussian particle filter (GPF) \cite{Kotecha03b}. This proposed algorithm contains the GPF as a special case (with $M=1$) and the regularized particle filter (with $M=N$) \cite{SMC01}.  The second proposed scheme, called {\it compressed particle filter} (C-PF), requires the evaluation of the measurement model only $M$ times instead of $N$. Therefore, the C-PF is faster than a standard particle filter and is particularly convenient when the likelihood evaluation is costly. We also provide an example of C-MC in modern adaptive IS schemes {to} allow the use of expensive mixtures as {the denominator} of the importance weights \cite{LAIS17,Ingmar}. More details  are provided in Section \ref{DistrSect}. 
Finally, note that similar and related ideas {have} been presented in different works and several applications, such as diffusion estimation  \cite{Chao15,Oreshkin10}, smoothing techniques \cite{Klaas06}, and as alternative resampling procedures in particle filtering \cite{Li15,Li_2012}. The benefits of the proposed schemes are shown in six different numerical experiments.
 }
 \newline 
 {\bf Structure of the work.} Section \ref{MC_Sect} introduces the basic setup of the Bayesian inference problem and describes the goal of the paper jointly with some possible solutions already presented in the literature. In Section \ref{SectCMC}, we introduce the C-MC method whereas, in Section \ref{AnC-MC}, we provide further analyses.  In Section \ref{DistrSect}, we describe different applications of C-MC, several novel algorithms, and further extensions.
Section \ref{Simu} provides six numerical experiments, and some conclusions are contained in Section \ref{ConSect}. The main acronyms of the work are summarized in Table \ref{table_acro}.

\begin{table}[!h]
\centering

\caption{\normalsize Main acronyms of the work.}
\vspace{-0.2cm}
	\begin{tabular}{|c|c|}
    \hline
pdf & probability density function\\
MC & Monte Carlo\\
QMC & Quasi-Monte Carlo \\
MCMC & Markov Chain Monte Carlo \\ 
IS & Importance Sampling \\ 
C-MC & Compressed Monte  Carlo\\ 
C-PF & Compressed Particle Filter\\ 
{MSE} & {Mean Square Error}\\ 
   \hline 
\end{tabular}
\label{table_acro}
\end{table}

\section{Background}
\label{MC_Sect}
\subsection{Problem statement}
In many real-world applications, the interest lies in obtaining information about the posterior density of {a set} of unknown parameters given the observed data.
Mathematically, denoting the vector of unknowns as  ${\bf x}=[x_1,...,x_{d_x}]^{\top}\in \mathcal{D}\subseteq \mathbb{R}^{d_X}$ and the observed data as ${\bf y}\in \mathbb{R}^{d_Y}$, the pdf is defined as
\begin{equation}
	{\bar \pi}({\bf x}| {\bf y})
		= \frac{\ell({\bf y}|{\bf x}) g({\bf x})}{Z({\bf y})} \propto \pi({\bf x}|{\bf y})=\ell({\bf y}|{\bf x}) g({\bf x}),
\label{eq_posterior}
\end{equation}
where $\ell({\bf y}|{\bf x})$ is the likelihood function, $g({\bf x})$ is the prior pdf, and $Z({\bf y})$ is the normalization factor, that is usually called marginal likelihood or Bayesian {model} evidence.
From now on, we remove the dependence on ${\bf y}$ to simplify the notation.
%
A particular integral involving the random variable ${\bf X} \sim {\bar \pi}({\bf x})=\frac{1}{Z} \pi({\bf x})$ is then given by
\begin{equation}
	I{(h)} \triangleq E_{\bar \pi}[h({\bf X})]= \int_{\mathcal{D}} h({\bf x}) \bar{\pi}({\bf x}) d{\bf x}= \frac{1}{Z} \int_{\mathcal{D}} h({\bf x}) \pi({\bf x}) d{\bf x},
\label{eq_integral}
\end{equation}
where $h({\bf x})$ can be any integrable function of ${\bf x}$.\footnote{{ To simplify the notation}, we have assumed $h({\bf x}):\mathbb{R}^{d_X}\rightarrow \mathbb{R}$ and the integral $I(h)\in \mathbb{R}$ is a scalar value. However,  a more proper assumption is ${\bf h}({\bf x}):\mathbb{R}^{d_X}\rightarrow \mathbb{R}^{\nu}$ and ${\bf I}({\bf h})\in \mathbb{R}^{\nu}$ where $\nu\geq1$. All the techniques and results in this work are valid for the more general mapping with $\nu\geq1$, but  we keep the simpler notation for $\nu=1$. With $\nu >1$, we would have a vector of integrals ${\bf I}({\bf h})$. For instance, if ${\bf h}({\bf x})={\bf x}$ we have $\nu=d_X$, and we have one integral for each component of ${\bf x}$.} {For simplicity,} we assume that the functions $h({\bf x})$ and $\bar{\pi}({\bf x})$ are continuous in $\mathcal{D}$, and the integrand function, $h({\bf x}) \bar{\pi}({\bf x})$, in Eq. \eqref{eq_integral} is integrable. More generally, we are interested in finding a particle approximation ${\widehat \pi}^{(N)}({\bf x})$ of the measure of ${\bar \pi}({\bf x})$ \cite{Liu04b}.
In many practical scenarios, we cannot obtain an analytical solution {for the integral in Eq.} \eqref{eq_integral}.
One possible alternative is to use different deterministic quadrature rules or formulas based on sigma points for approximating the integral $I(h)$ \cite{CKF09,Julier04,Sarkka13bo}. However, these  deterministic techniques are available only in specific scenarios, i.e., for some particular pdfs $\bar{\pi}({\bf x})$. Hence, Monte Carlo schemes are often preferred and applied {to} estimate $I$ and provide a particle approximation ${\widehat \pi}^{(N)}({\bf x})$.
\subsection{Monte Carlo {(MC)} sampling techniques}
If it is possible to draw $N$ independent samples, $\{{\bf x}_n\}_{n=1}^N$, directly from $\bar{\pi}({\bf x})$, then we can construct a particle approximation ${\widehat \pi}^{(N)}({\bf x})= \frac{1}{N}\sum_{n=1}^N  \delta({\bf x}-{\bf x}_n)$ of the measure of ${\bar \pi}$ \cite{Robert04}. { This is the foundation of MC methods, denote as standard or direct MC.} Therefore, replacing ${\bar \pi}({\bf x})$ with ${\widehat \pi}^{(N)}({\bf x})$ in Eq. \eqref{eq_integral}, we obtain the standard Monte Carlo estimator of $I$,
\begin{equation}
\label{S_MC_MCMC_est}
\widehat{I}^{(N)}{(h)} = \frac{1}{N}\sum_{n=1}^N  h({\bf x}_n).
\end{equation}
{ However,} when sampling from $\bar{\pi}({\bf x})$ is not possible, alternative MC methods are used \cite{Liu04b,Robert04}{.} For instance, the MCMC algorithms generate correlated samples $\{{\bf x}_n\}_{n=1}^N$ that, after a burn-in period, are distributed according to $\bar{\pi}({\bf x})$. 
Another possible approach is based on the importance sampling (IS) technique \cite{Robert04,7974876}. { In the following, we describe the basic ideas behind the IS schemes. Consider} $N$ samples $\{{\bf x}_n\}_{n=1}^N$ drawn from a proposal pdf, $q({\bf x})$, with heavier tails than the target, ${\bar \pi}({\bf x})$. We assign a weight to each sample and then we can be normalized them as follows,
\begin{equation} 
	w_i= \frac{\pi({\bf x}_i)}{{q({\bf x}_i)}}, \qquad  \qquad \bar{w}_i=\frac{w_i}{\sum_{j=1}^N w_j},
\label{is_weights_static}
\end{equation} 
with $i=1,...,N$. Therefore, the moment of interest can be approximated as
\begin{eqnarray}
	\widehat{I}^{(N)}{(h)}&=& \frac{1}{N\widehat{Z}}  \sum_{i=1}^N w_i h({\bf x}_i)  \\
	&=& \sum_{i=1}^N \bar{w}_i h({\bf x}_i) ,
\label{eq_partial_estimator_static}
\end{eqnarray}
where $\widehat{Z}=\frac{1}{N}\sum_{j=1}^N w_j$ is a unbiased estimator of $Z=\int_{\mathcal{D}} \pi({\bf x}) d{\bf x}$ \cite{Robert04}. 
{ One can consider that, in the standard Monte Carlo and MCMC methods, the normalized weights are $\bar{w}_i=1/N$.
Then, all the described Monte Carlo estimators can be summarized by Eq. \eqref{eq_partial_estimator_static}, and the particle approximation of the measure of ${\bar \pi}$ is given by
\begin{equation}
\label{EqPiMonteCarlo}
{\widehat \pi}^{(N)}({\bf x})= \sum_{n=1}^N \bar{w}_n \delta({\bf x}-{\bf x}_n),
\end{equation}
where $\delta({\bf x})$ is the Dirac delta function. This formulation encompasses jointly {MCMC and IS}, and in the former case, we have access to the values of the unnormalized weights $w_n$. Hence, {in the IS setting}, an estimator $\widehat{Z}=\frac{1}{N}\sum_{n=1}^N w_n$ of the marginal likelihood $Z$ is also available. 
}

\subsection{Goal}
\label{GoalSectv0}

In this work, we address the problem of summarizing the information contained in a set of $N$ weighted or unweighted samples generated by a Monte Carlo sampling technique, with a smaller amount $M<N$ of  weighted samples. This problem is strictly related to the more general challenge: summarizing the required information of a given target density ${\bar \pi}({\bf x})$, using a particle approximation (with the smallest amount of weighted particles). {Generally,} there is a loss of information.  More precisely, given a Monte Carlo approximation ${\widehat \pi}^{(N)}({\bf x})$ in Eq. \eqref{EqPiMonteCarlo}, with $N$ samples, we desire to construct another particle approximation
\begin{equation}
{\widetilde \pi}^{(M)}({\bf x})= \sum_{m=1}^M \bar{a}_m \delta({\bf x}-{\bf s}_m),
\end{equation}
where $M<N$, $\sum_{m=1}^M  \bar{a}_m =1$, and ${\bf s}_m \in \mathcal{D}$, sharing with ${\widehat \pi}^{(N)}$ the required properties. The goal is to compress the statistical information contained in ${\widehat \pi}^{(N)}({\bf x})$, reducing as much as possible the loss of information. We refer to $ \bar{a}_m$ as summary weights and, to ${\bf s}_m$, as summary particles. The rate of compression is clearly given by $\eta=\frac{N}{M}$. Note that when $\eta=1$ we have  no compression whereas, when $\eta=N$, we have the maximum compression ($1 \leq \eta \leq N$). 

\subsection{Related works}
\label{GoalSect}
In the literature, two families of possible solutions have been proposed for different but related purposes.  
The first one is based on a bootstrap technique, and can be always used. The second one is the moment-matching approach, and is available only for a limited type target pdfs ${\bar \pi}({\bf x})$. 
\newline
\newline
{\bf Bootstrap solution.} {Let assume that} we have $N$ unweighted samples. A simple approach for compression consists in choosing uniformly $M$ samples within the $N$ possible ones. Similarly, in the case of weighted samples, this strategy consists in resampling $M$ times within the set $\{{\bf x}_n\}_{n=1}^N$ according to the normalized  weights $\bar{w}_n$, $n=1,...,N$ \cite{Bolic05}. Then, a proper aggregated weight is associated to the resampled particles \cite{Bolic05,GISssp16,GIS18}. This kind of compression scheme has been widely used in different works (explicitly or implicitly), from distributed particle filtering methods and other sophisticated Monte Carlo algorithms \cite{Bolic05,Read2014,Miguez16,Pisland15}.
\newline
\newline 
{\bf Moment-matching solution.} For simplicity and without loss of generality, let us consider $d_X=1$, i.e., $x\in  \mathbb{R}$. For some specific types of target pdfs ${\bar \pi}(x)$ and specific domains $\mathcal{D}$, it is possible to obtain a deterministic particle approximation 
${\widetilde \pi}^{(M)}({\bf x})= \sum_{m=1}^M \rho_m \delta( x-s_m) $ where the weights $\rho_m$ and the particles $s_m$ are solutions of the nonlinear moment-matching system below,
\begin{eqnarray}
\label{SystemOpt}
\sum_{m=1}^M \rho_m  s_m^r=\int_{\mathcal{D}} x^r{\bar \pi}(x) dx  \quad \mbox{ for } \quad r=1,...,R=2M,
\end{eqnarray}
where the true values of the first $2M$ non-central moments, $\int_{\mathcal{D}} x^r{\bar \pi}(x) dx$, must be known. Hence we have $2M$ unknowns (the $M$ weights $\rho_m$ and the $M$ particles $s_m$) and $R=2M$ equations. Since the system is highly nonlinear, in general, the analytical solution is available only in few particular cases. These solutions are {known as} {\it Gaussian Quadratures} \cite{Sarkka13bo}, the corresponding deterministic particle approximation provide a perfect-matching with the first $2M$ moments (zero loss of information in the approximation of these moments). Quadrature rules and related sigma point methods have been widely applied within several generalized Kalman filtering techniques \cite{CKF09,Julier04,Sarkka13bo}.  



\begin{figure*}[htbp]
\centering
\includegraphics[width=8cm]{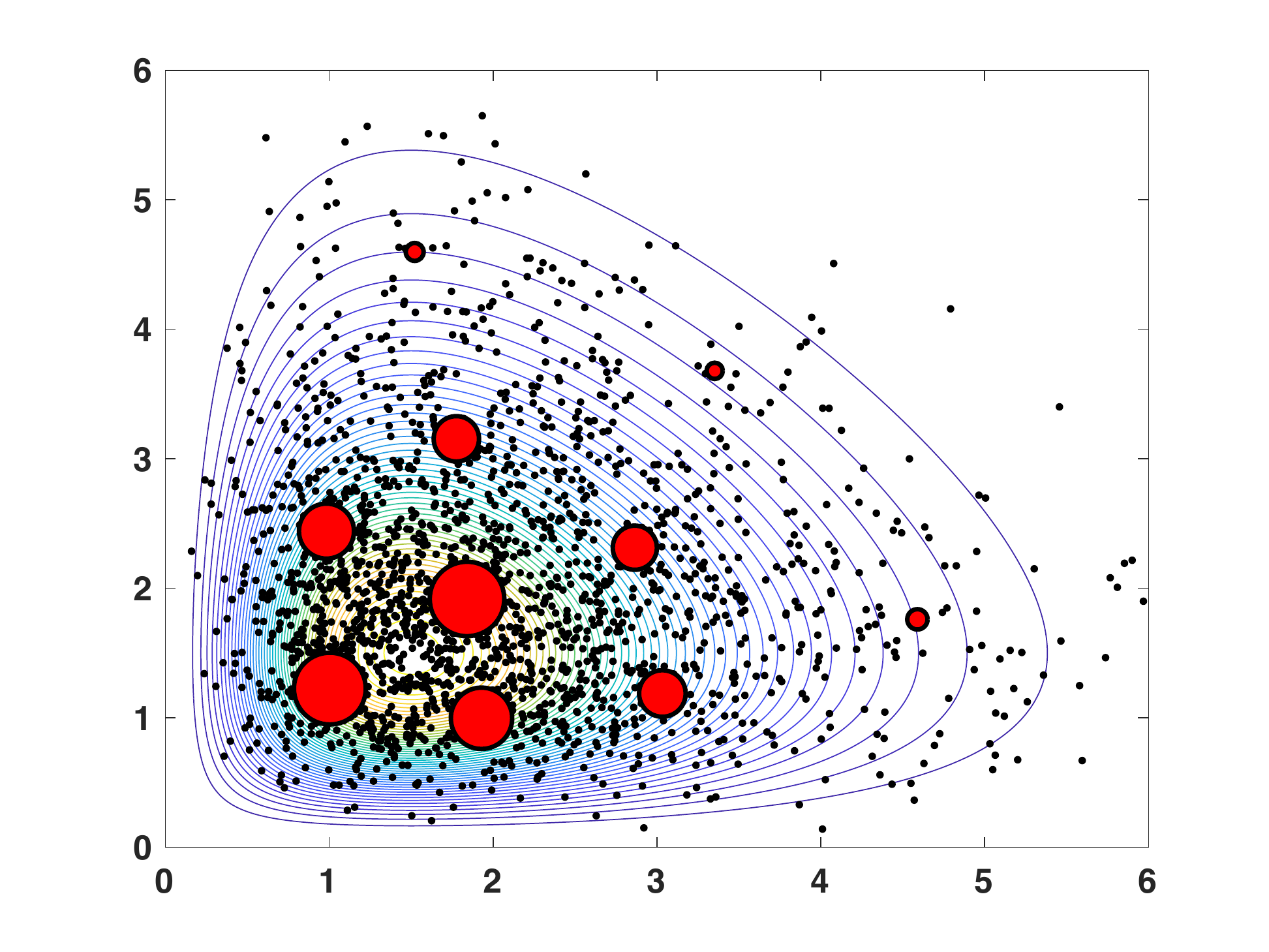}
\includegraphics[width=8cm]{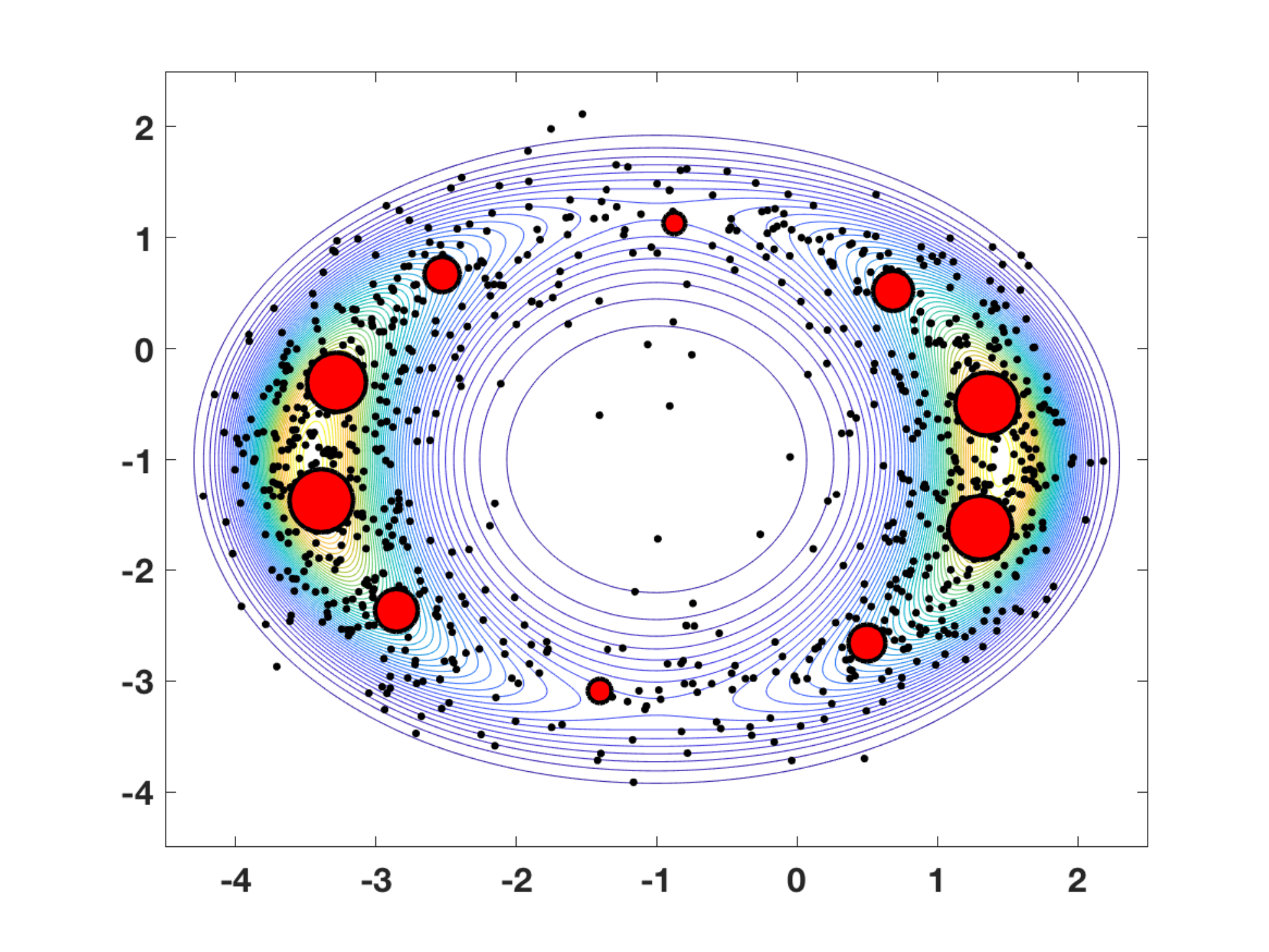}
\caption{One run of a C-MC scheme with $M=10$, for two different clouds of $N=10^3$ samples (represented by dots). Each figure represents a different target ${\bar \pi}({\bf x})$ (shown by the contour plots). The size of the circles is proportional to the corresponding summary weight. }
\label{Fig2}
\end{figure*}




\section{Compressed Monte Carlo (C-MC)}
\label{SectCMC}

 In this work, we introduce a compression approach {that improves} the bootstrap strategy and extends the applicability of the moment-matching scheme, both described above. We consider the cases of compressing unweighted and weighted samples, e.g., the $N$ samples have been previously generated by an MCMC algorithm or an IS technique, respectively.  Figure \ref{Fig2} shows two examples of C-MC approximation with $M=10$ summary particles. The size of the circles is proportional to the corresponding summary weight.   

\subsection{Stratification}
\label{StrataSect}

The underlying grounds of C-MC are based on the so-called stratified sampling \cite{Ecuyer94,mcbookOwen}. The idea is to divide the support domain $\mathcal{D}$ of the random variable ${\bf X}$ into $M$ separate and mutually
exclusive regions.  More specifically, let us consider an integer $M \in \mathbb{N}^+$, and a partition $\mathcal{P}=\{\mathcal{X}_1,\mathcal{X}_2, ....,\mathcal{X}_M\}$ of the state space with $M$ disjoint subsets, 
\begin{gather}
\begin{split}
\label{PartitionEq}
&\mathcal{X}_1\cup\mathcal{X}_2\cup...\cup\mathcal{X}_M=\mathcal{D},\\
&\mathcal{X}_i\cap\mathcal{X}_k =\emptyset, \qquad i\neq k, \quad \forall i,j\in\{1,...,M\}. 
\end{split}
\end{gather}
 We assume that all $\mathcal{X}_m$ are convex sets.
Then, in the simplest version {of the stratification approach}, one sample is drawn from each sub-region, and finally all the generated samples are combined for providing an estimator of $I(h)$. We also denote the area of ${\bar \pi}({\bf x})$ restricted in $\mathcal{X}_m$ as
\begin{gather}
\label{amEq}
\begin{split}
{\bar a}_m =\mathbb{P}({\bf X}\in \mathcal{X}_m) &= \int_{\mathcal{X}_m} {\bar \pi}({\bf x}) d{\bf x}=\frac{1}{Z} \int_{\mathcal{X}_m}  \pi({\bf x}) d{\bf x},   \\
&=\frac{Z_m}{Z}=\frac{Z_m}{\sum_{j=1}^M Z_j}, 
\end{split}
\end{gather}
where $Z_m=\int_{\mathcal{X}_m}  \pi({\bf x}) d{\bf x}$ and  $Z=\sum_{j=1}^M Z_j=\int_{\mathcal{D}} \pi({\bf x}) d{\bf x}$. {Note} that $\sum_{m=1}^M {\bar a}_m=1$. The target density can be expressed as a mixture of $M$ non-overlapped densities,
\begin{eqnarray}
{\bar \pi}({\bf x})=\sum_{m=1}^M  {\bar a}_m \left[\frac{1}{{\bar a}_m}  \bar{\pi}({\bf x}) \mathbb{I}_{\mathcal{X}_m}({\bf x})\right]= \sum_{m=1}^M  {\bar a}_m {\bar \pi}_m({\bf x}), 
\end{eqnarray}
where 
\begin{eqnarray}
{\bar \pi}_m({\bf x})=\frac{1}{{\bar a}_m} {\bar \pi}({\bf x}) \mathbb{I}_{\mathcal{X}_m}({\bf x})=\frac{1}{Z_m} \pi({\bf x}) \mathbb{I}_{\mathcal{X}_m}({\bf x}),
\end{eqnarray}
 is {the $m$-th} density {in the mixture}, and $\mathbb{I}_{\mathcal{X}_m}({\bf x})$ is an indicator function that is $1$ when ${\bf x}\in \mathcal{X}_m$ and $0$ otherwise. 
\newline
\newline
{\bf Stratified MC estimators.} In order to simulate a sample ${\bf x}^*$ from ${\bar \pi}({\bf x})$, we can draw an index $j^*\in\{1,...,M\}$ according to the probability mass function $\bar{a}_m$, $m=1,...,M$ and the draw ${\bf x}^* \sim {\bar \pi}_{j^*}({\bf x})$. Alternatively, we {could} yield an approximation of the measure of ${\bar \pi}$, drawing one sample from each region, i.e., ${\bf s}_m \sim {\bar \pi}_{m}({\bf x})$, and then assign to each sample the weight $\bar{a}_m$, $m=1,...,M$. Therefore, in this scenario, the corresponding estimator of the integral $I(h)$ in Eq. \eqref{eq_integral} and the particle approximation are, respectively,
\begin{eqnarray}
\label{Strat_I_Eq}
&&\widetilde{I}^{(M)}(h)= \sum_{m=1}^M \bar{a}_m  h({\bf s}_m),  \mbox{ and } \\
&& {\widetilde \pi}^{(M)}({\bf x})= \sum_{m=1}^M \bar{a}_m \delta({\bf x}-{\bf s}_m),
\end{eqnarray}
where 
${\bf s}_m \sim {\bar \pi}_{m}({\bf x})=\frac{1}{Z_m} {\pi}({\bf x}) \mathbb{I}_{\mathcal{X}_m}({\bf x})$,
hence ${\bf s}_m  \in \mathcal{X}_m$. See the \texttt{Supplementary Material}, for extensions and further details. 

\subsection{C-MC algorithms}
{Let consider} $N$ weighted samples $\{{\bf x}_n,{\bar w}_n\}_{n=1}^N$ generated by a MC scheme, and {let $M$ be a constant value such that $M<N$}. Given the partition in Eq. \eqref{PartitionEq}, i.e., $\mathcal{X}_1\cup\mathcal{X}_2\cup...\cup\mathcal{X}_M=\mathcal{D}$ formed by convex, disjoint sub-regions $\mathcal{X}_m${, we} denote the subset of the set  of indices $\{1,...,N\}$, 
$$
\mathcal{J}_m=\{i=1,...,N: \mbox{ } {\bf x}_i \in \mathcal{X}_m\},
$$
which are associated with the samples in the $m$-th sub-region $\mathcal{X}_m$. The cardinality $|\mathcal{J}_m|$ denotes the number of samples in $\mathcal{X}_m$, and we have $\sum_{m=1}^{M} |\mathcal{J}_m|=N$.
\newline 
\newline
{\bf C-MC approximation.} We can compress the information contained in the particle approximation  
of Eq. \eqref{EqPiMonteCarlo}, constructing an empirical stratified approximation based on $M$ weighted particles $\{{\bf s}_m,\widehat{a}_m\}_{m=1}^M$, i.e.,
\begin{eqnarray}  
\label{CMC_PI_Eq}
{\widetilde \pi}^{(M)}({\bf x})= \sum_{m=1}^M \widehat{a}_m \delta({\bf x}-{\bf s}_m),
\end{eqnarray}
so that for a specific moment the resulting estimator is 
\begin{eqnarray}
\label{CMC_I_Eq}
{\widetilde I}{^{(M)}(h)}= \sum_{m=1}^M \widehat{a}_m h({\bf s}_m),
\end{eqnarray}
where $\widehat{a}_m$ is an approximation of  ${\bar a}_m = \int_{\mathcal{X}_m} {\bar \pi}({\bf x}) d{\bf x}$  in Eq. \eqref{amEq}, {considering the given samples.}
\newline 
\newline
{\bf Normalized C-MC weights.} We can write
\begin{eqnarray}
\widehat{a}_m&=&\int_{\mathcal{X}_m} \widehat{\pi}^{(N)}({\bf x}) d{\bf x}= \sum_{i=1}^N {\bar w}_i \int_{\mathcal{X}_m} \delta({\bf x}-{\bf x}_i) d{\bf x}, \nonumber \\
&=&  \sum_{i\in \mathcal{J}_m} {\bar w}_i.
\end{eqnarray}
Hence, in the case of compressing samples generated by a standard MC or MCMC schemes, since ${\bar w}_i=\frac{1}{N}$, we obtain $\widehat{a}_m=\frac{| \mathcal{J}_m |}{N}$ that is {again} an estimate of the probability ${\bar a}_m =\mathbb{P}({\bf X}\in \mathcal{X}_m)$ {in Eq. \eqref{amEq}}. In the IS case, we can also obtain the expression $\widehat{a}_m$ as the ratio of the MC estimators 
\begin{eqnarray}
\widehat{Z}_m=\frac{1}{N} \sum_{i\in \mathcal{J}_m} w_i, \qquad  \widehat{Z}= \sum_{m=1}^M \widehat{Z}_m=\frac{1}{N} \sum_{n=1}^N w_n,
\end{eqnarray}
i.e., 
\begin{eqnarray}\label{EqAmSum}
\widehat{a}_m=\frac{\widehat{Z}_m}{\widehat{Z}}
=\sum_{i\in \mathcal{J}_m} \frac{w_i}{\sum_{n=1}^N w_n}= \sum_{i\in \mathcal{J}_m} {\bar w}_i,
\end{eqnarray}
as suggested by Eq. \eqref{amEq}. {Note that, in all cases, we have $0 \leq \widehat{a}_m\leq 1$ and $
\sum_{m=1}^M \widehat{a}_m=1$.}
\newline 
\newline
{{\bf Stochastic choice of ${\bf s}_m$.} We consider different strategies for the selection of the summary particles ${\bf s}_m$. The first one is a stochastic approach based on the stratified sampling: each summary particle ${\bf s}_m$ is resampled within the set  of samples ${\bf x}_i \in \mathcal{X}_m$, i.e., $\{{\bf x}_i, \mbox{ with } i \in \mathcal{J}_m\}$, according to the normalized weights,
\begin{equation}
\label{EqREVmagic}
{\bar w}_{m,i}=\frac{w_i}{\sum_{k\in \mathcal{J}_m} w_k}=\frac{\bar{w}_i}{\sum_{k\in \mathcal{J}_m}\bar{w}_k}=\frac{\bar{w}_i}{\widehat{a}_m}, \quad i \in \mathcal{J}_m.
\end{equation}
{In the case} of samples {generated} by standard MC or MCMC schemes, {then we obtain} ${\bar w}_{m,i}=\frac{1}{|\mathcal{J}_m|}$.
 }
\newline 
\newline
{{\bf Deterministic choice of ${\bf s}_m$.} In the same fashion of the deterministic rules and sigma-point construction discussed in Section \ref{GoalSect}, we can also set 
\begin{equation}
\label{VDet}
{\bf s}_m=\sum_{j\in \mathcal{J}_m}  {\bar w}_{m,j} {\bf x}_{j}, 
\end{equation}
or, if we are interested on the approximation of a specific integral involving a function $h$, we can set
\begin{equation}
\label{EqSpecH}
s_m=\sum_{j\in \mathcal{J}_m}   {\bar w}_{m,j} h({\bf x}_{j}). 
\end{equation}
These deterministic rules provides a good performance and enjoy interesting properties, {as discussed} in the next sections {and Appendix \ref{App1}}.
}
\newline
\newline
{ {\bf Other C-MC weights.}  {In some applications,} it is required to define an {\it aggregated weight} 
\begin{equation}
W=\sum_{n=1}^N w_n=N \widehat{Z},
\end{equation}
which is associated to the discrete measure ${\widetilde \pi}^{(M)}$. 
 It is useful { in the distributed scenario, as described in Section \ref{DistrSect}} \cite{Bolic05,Read2014,GIS18}. In the case of samples {drawn} by a standard Monte Carlo or MCMC scheme, the unnormalized weights $w_n$ are unknown{, but} we can set $W=N$. In the IS scenario, we can also define the unnormalized C-MC weights $a_m=\frac{1}{N}\widehat{a}_m W=\widehat{Z}_m$.  These weights, $a_m$, can be employed for reconstructing the estimator of {the} marginal likelihood. Indeed, we have
 \begin{gather}
 \label{Zcis}
 \begin{split}
\widetilde{Z}&=\frac{1}{N}\sum_{m=1}^M a_m=\frac{1}{N}\sum_{m=1}^M \widehat{Z}_m=\widehat{Z},
\end{split}
\end{gather}
recovering perfectly the IS estimator $\widehat{Z}$. Table \ref{CMCsumm} summarizes the main expressions {introduced} in this section.
}

 \begin{table}[!h]
\small
{ 
\caption{Summary of the main  C-MC expressions.}
\label{CMCsumm}
\begin{tabular}{|c|c|c||c|c|c|c|}
   \hline
 Scheme & $w_i$ & ${\bar w}_i$ &  \multicolumn{2}{c|}{$\widehat{a}_m$}   &  $a_m$   & $W$ \\
\hline
\hline
      & &    &       & &   & \\ 
IS & $\frac{\pi({\bf x}_i)}{{q({\bf x}_i)}}$ & $\frac{w_i}{\sum\limits_{n=1}^N w_n}$  & \multirow{6}{*}{$\sum\limits_{i\in \mathcal{J}_m} {\bar w}_i$} &  $\dfrac{\widehat{Z}_m}{\widehat{Z}}$   &  $\widehat{Z}_m$ &  $\sum\limits_{i=1}^N w_i$ \\
& & &  & && \\
\cline{1-3}
\cline{5-7}
   & &   &    & &   & \\
  MCMC & ---  &  $\dfrac{1}{N}$ & & $\dfrac{|\mathcal{J}_m|}{N}$  & --- & $N$\\
 & &  &   &  & & \\
\hline
\multicolumn{7}{|c|}{ } \\
 \multicolumn{7}{|c|}{$\widehat{Z}_m =\frac{1}{N}\sum_{i\in \mathcal{J}_m} w_i \qquad  \widehat{Z} =\frac{1}{N}\sum_{i=1}^N w_i \qquad {\bar w}_{m,j}=\dfrac{{\bar w}_i}{\widehat{a}_m}$ } \\ 
\multicolumn{7}{|c|}{ } \\
\hline
\end{tabular}
}
\end{table}

\noindent 
{ {\it Additional observation.} Note also that {the estimator ${\widehat I}^{(N)}(h)$ can be expressed as linear combination of partial estimators, i.e.,}  
\begin{eqnarray}
{\widehat I}^{(N)}(h)=\sum_{i=1}^N {\bar w}_i h({\bf x}_i)&=&\sum_{m=1}^M \sum_{i\in\mathcal{J}_m} {\bar w}_i h({\bf x}_i),  \nonumber \\
&=&\sum_{m=1}^M \widehat{a}_m \sum_{i\in\mathcal{J}_m} {\bar w}_{m,i} h({\bf x}_i) \nonumber\\
&=&\sum_{m=1}^M \widehat{a}_m {\widehat I}_m(h),  \label{aquiComb_I}
\end{eqnarray}
where  we have used ${\bar w}_{m,i}= \frac{{\bar w}_i}{\widehat{a}_m}$ as shown in Eq. \eqref{EqREVmagic}, and we have define {the partial estimators ${\widehat I}_m(h)=\sum_{i\in\mathcal{J}_m} {\bar w}_{m,i} h({\bf x}_i)$}, for $m=1,...,M$. Namely, the MC estimator ${\widehat I}^{(N)}(h)$ of $I(h)$ can be expressed as a {{\it convex} combination of the $M$  partial MC estimators, since $
\sum_{m=1}^M \widehat{a}_m=1$.} A similar expression is valid for the particle approximations, i.e.,
\begin{eqnarray}
&&{\widehat \pi}^{(N)}({\bf x})=\sum_{m=1}^M \widehat{a}_m {\widehat \pi}_m({\bf x}), \quad \mbox{where}  \\
&&{\widehat \pi}_m({\bf x})=\sum_{i\in\mathcal{J}_m} {\bar w}_{m,i} \delta ({\bf x}-{\bf x}_i).
\end{eqnarray}
}


\section{Analysis of C-MC}
\label{AnC-MC}
\noindent
{\it Proper partition and consistency.} Let us focus {on} the way the partition is formed.
A partition rule is proper if, when $M=N$, then $|\mathcal{J}_m|=1$ (note that $m=n$ in this case), i.e., in the limit case of $M=N$ we consider all the MC samples as summary samples. Recall that, for $M<N$, the C-MC estimators are unbiased as shown in the \texttt{Supp. Material} (with $K_m=1$ and $V=M$).  
Furthermore, if the partition rule is proper { then, for $M=N$, the} C-MC estimators {will} coincide with the {non-compressed MC} estimators. Hence, as $M\rightarrow N$ and $N\rightarrow \infty$, the consistency is ensured.
\newline
\newline
{\it Save in transmission.} Let us consider the parallel or distributed framework with a common central node. In C-MC, only the $M$ pairs $\{\widehat{a}_m,{\bf s}_m\}_{m=1}^M$ are transmitted to the central node, instead of the $N$ pairs. Since, ${\bf x},{\bf s}\in \mathbb{R}^{d_X}$, without compression, we need to transmit $Nd_X$ scalar values in case {of} unweighted samples, or $N(d_X+1)$ scalar values in the case of weighted samples.
With the proposed compression {scheme}, the transmission of only $M(d_X+1)$ scalar values {is} required.

\subsection{Compression Loss}
\label{LossSect}
{
\noindent
{\it Loss for the deterministic C-MC.} Let us consider the deterministic choice of the summary particles as
\begin{equation}
\label{Sm_aqui}
{\bf s}_m=\sum_{j\in \mathcal{J}_m}  {\bar w}_{m,j} {\bf x}_{j}, \qquad m=1,...,M.
\end{equation}
Hence, { keeping fixed $\{{\bf x}_{n},{\bar w}_n\}_{n=1}^N$ and the partition}, the summary particles ${\bf s}_m$ {defined in Eq. \eqref{Sm_aqui}} are {also} fixed.
Recall that the standard MC estimator and the corresponding C-MC estimator are
\begin{eqnarray*}
\widehat{I}^{(N)}(h)=  \sum_{n=1}^N  \bar{w}_n h({\bf x}_n), \quad {\widetilde{I}^{(M)}}(h)=  \sum_{m=1}^M  \widehat{a}_m h({\bf s}_m).
\end{eqnarray*}
For a specific function $h$, the information loss for a C-MC scheme can be measured with the squared error, i.e., 
\begin{eqnarray}
\ell(h)=(\widehat{I}^{(N)}(h)-{\widetilde{I}^{(M)}}(h))^2,
\end{eqnarray}
or more generally,
\begin{eqnarray}
\ell(h,f)=(\widehat{I}^{(N)}(h)-{\widetilde{I}^{(M)}}(f))^2,
\end{eqnarray}
where $f({\bf x})$ is another integrable function. Furthermore, considering a family $\mathcal{H}$ of $R$ functions, i.e., $\mathcal{H}=\{h_1({\bf x}),..., h_R({\bf x})\}$, we can write  
 we can define the loss as 
\begin{eqnarray}
\label{LOSS}
\mathcal{L}_R=\sum_{r=1}^R \xi_r^2 \ell(h_r)=\sum_{r=1}^R  \xi_r^2 \left(\widehat{I}^{(N)}(h_r)-{\widetilde{I}^{(M)}}(h_r)\right)^2,
\end{eqnarray}
which is a weighted average of the squared errors, with weights $\xi_r^2$. For instance, we can set $\xi_r^2\propto \frac{1}{\left[\widehat{I}^{(N)}(h_r)\right]^2}$ if $\widehat{I}^{(N)}(h_r) \neq 0${,} so that $\mathcal{L}_R$ is equivalent to a sum of the relative errors, or simply $\xi_r^2=\frac{1}{R}$.
Moreover, recalling that ${\widehat I}^{(N)}(h)=\sum_{m=1}^M \widehat{a}_m {\widehat I}_m(h)$ as shown in Eq. \eqref{aquiComb_I}, we can write 
\begin{eqnarray}
 \ell(h) &=& \left(\widehat{I}^{(N)}(h)-{\widetilde{I}^{(M)}}(h)\right)^2  \nonumber \\
 &=&\left(\sum_{m=1}^M  \widehat{a}_m  \sum_{j\in \mathcal{J}_m}  \bar{w}_{m,j}h({\bf x}_j)-\sum_{m=1}^M \widehat{a}_m h({\bf s}_m)\right)^2.  \nonumber 
  \end{eqnarray}
 {We  can rewrite it as}
 \begin{eqnarray} 
 \ell(h) =\left(\sum_{m=1}^M \widehat{a}_m \left[ \sum_{j\in \mathcal{J}_m}  \bar{w}_{m,j}h({\bf x}_j)-h({\bf s}_m)\right]\right)^2.    \nonumber 
 \end{eqnarray}
Recalling $\widehat{\pi}_{m}({\bf x})=\sum\limits_{j\in \mathcal{J}_m}  {\bar w}_{m,j} \delta({\bf x}-{\bf x}_{j})$ and the definition of ${\bf s}_m$ in Eq. \eqref{Sm_aqui}, we can also write
\begin{eqnarray}
  \ell(h)
 &=&\left(\sum_{m=1}^M c_m(h)\right)^2, \label{L1}
\end{eqnarray}
{where
\begin{eqnarray}
c_m(h)
= \widehat{a}_m   \Big[ \sum_{j\in \mathcal{J}_m}  \bar{w}_{m,j}h({\bf x}_j)-h\Big(\sum_{j\in \mathcal{J}_m}  {\bar w}_{m,j} {\bf x}_{j}\Big)\Big],   \label{cmh_aqui} 
\end{eqnarray}
and we have replaced the specific choice ${\bf s}_m=\sum_{j\in \mathcal{J}_m}  {\bar w}_{m,j} {\bf x}_{j}$ in Eq. \eqref{Sm_aqui}.}
Using {also the equalities} $\bar{w}_{m,j}=\frac{\bar{w}_j}{\widehat{a}_m}$ and $\widehat{a}_m=\sum_{j\in \mathcal{J}_m}  {\bar w}_{j}$, {we obtain}
\begin{eqnarray}
c_m(h)&=&\sum_{j\in \mathcal{J}_m}  \bar{w}_jh({\bf x}_j)- \widehat{a}_m h\Big(\sum_{j\in \mathcal{J}_m}  {\bar w}_{m,j} {\bf x}_{j}\Big), \\
&=&\sum_{j\in \mathcal{J}_m}  \bar{w}_j \Big[ h({\bf x}_j)-  h\Big(\sum_{j\in \mathcal{J}_m}  {\bar w}_{m,j} {\bf x}_{j}\Big)\Big]. \label{C1}
\end{eqnarray}
 The expressions \eqref{cmh_aqui}-\eqref{C1} only depend on the MC samples $\{{\bf x}_n, {\bar w}_n\}$ and the partition{, that we have considered pre-established and fixed.
Note also that if $h$ is a linear function, then we have a zero-loss compression, i.e., $\ell(h)=0$.} 
 The choice in Eq. \eqref{Sm_aqui} is interesting since it provides a very good performance (see Section \ref{Simu}) and also resembles a deterministic quadrature rule with weighted nodes (it can be interpreted an approximate sigma-point construction \cite{Julier04,Sarkka13bo}).  
 \newline
  \newline
{\bf Zero-loss compression.} If we are interested only in one specific integral $I{(h)}= \int_{\mathcal{D}} h({\bf x}) \bar{\pi}({\bf x}) d{\bf x}$, it is convenient to apply C-MC with the following summary particles  
\begin{equation}
\label{EqSpecH2}
s_m=\sum_{j\in \mathcal{J}_m} {\bar w}_{m,j} h({\bf x}_{j}),
\end{equation}
as highlighted by the theorem below.
{\theo\label{Teo1}  If  ${s}_m$ as in Eq. \eqref{EqSpecH2} is chosen, for $m=1,...,M$, and the linear mapping $f(x)=x$, we have $\widehat{I}^{(N)}(h)={\widetilde I}^{(M)}(f)$, i.e., zero-compression loss $\ell(h,f)=0$. 
} 
\newline
\newline
See Appendix \ref{App1} for the proof. Therefore, if we are interested only in one specific integral involving ${\bar \pi}({\bf x})$, we can obtain a perfect compression by choosing the summary particles as in Eq. \eqref{EqSpecH2}. { With the choice in Eq. \eqref{EqSpecH2}, $s_m\in \mathbb{R}$ is a scalar value since we have assumed $h({\bf x}):\mathbb{R}^{d_X}\rightarrow \mathbb{R}$ for simplicity, instead of the more general assumption ${\bf h}({\bf x}):\mathbb{R}^{d_X}\rightarrow \mathbb{R}^{s}$, and $s\geq1$. However, all the presented results are valid for the general case with $s\geq1$.}
 \newline
 \newline
{{\it  Zero-loss estimator of the marginal likelihood}}. In the weighted sample scenario, we have also the estimator of the marginal likelihood $\widehat{Z}=\frac{1}{N}\sum_{n=1}^N  w_n$. The corresponding C-MC estimator is ${\widetilde I}^{(M)}=\frac{1}{M} \sum_{m=1}^M a_m=\frac{1}{M} \sum_{m=1}^M Z_m =\widehat{Z}$ as shown in Eq. \eqref{Zcis}, hence the loss is $\left({\widetilde I}^{(M)}-\widehat{Z}\right)^2=0$. Namely, we always recover the IS estimator of the marginal likelihood, without any loss.
 \newline
 \newline
{\it Loss for the stochastic C-MC.}  Let us consider the case when ${\bf s}_m$ is resampled randomly in each partition, according to the weights $ {\bar w}_{m,j}$ in Eq. \eqref{EqREVmagic}. Given the set of weighted samples $\mathcal{S}=\{{\bf x}_n,{\bar w}_n\}_{n=1}^N$, we can define the conditional expected mean-square error,  
\begin{eqnarray}
\ell(h)=\mbox{E}_{\widehat{\pi}_m}[({\widetilde{I}^{(M)}}(h)-\widehat{I}^{(N)}(h))^2|\mathcal{S}].
\end{eqnarray}
Note that, in this case, {we have}  
\begin{eqnarray}
\mbox{E}_{\widehat{\pi}_m}[h({\bf s}_m)|\mathcal{S}]=\sum_{j\in \mathcal{J}_m} {\bar w}_{m,j} h({\bf x}_{j})=\widehat{I}_m(h).
\end{eqnarray}
Given Eq. \eqref{aquiComb_I}, we can also write {as} 
\begin{eqnarray}
\widehat{I}^{(N)}(h)=\sum_{m=1}^M \widehat{a}_m \widehat{I}_m(h)=\sum_{m=1}^M \widehat{a}_m \mbox{E}_{\widehat{\pi}_m}[h({\bf s}_m)|\mathcal{S}],
\end{eqnarray}
so that 
\begin{eqnarray*}
{\widetilde{I}^{(M)}}(h)-\widehat{I}^{(N)}(h)
=\sum_{m=1}^M \widehat{a}_mÊ\Big( h({\bf s}_m)- \mbox{E}_{\widehat{\pi}_m}[h({\bf s}_m)|\mathcal{S})]\Big).
\end{eqnarray*}
Taking the expectation of both sides, we have 
\begin{eqnarray*}
&&\mbox{E}_{\widehat{\pi}_m}[{\widetilde{I}^{(M)}}(h)-\widehat{I}^{(N)}(h)|\mathcal{S}] \\
&&=\sum_{m=1}^M \widehat{a}_m Ê\Big(Ê\mbox{E}_{\widehat{\pi}_m}[h({\bf s}_m|\mathcal{S}]- \mbox{E}_{\widehat{\pi}_m}[h({\bf s}_m)|\mathcal{S}]\Big)=0,
\end{eqnarray*}
where we have also used the property $\mbox{E}[\mbox{E}[Z]]=\mbox{E}[Z]$. Therefore, the conditional mean error is zero, then
conditional expected mean-square error can be easily expressed as
\begin{eqnarray}
\ell(h)&=&\mbox{E}_{\widehat{\pi}_m}[({\widetilde{I}^{(M)}}(h)-\widehat{I}^{(N)}(h))^2|\mathcal{S}] \nonumber \\
&=&\sum_{m=1}^M\widehat{a}_m^2Ê\mbox{E}_{\widehat{\pi}_m}\Big[ \Big( h({\bf s}_m)- \mbox{E}_{\widehat{\pi}_m}[h({\bf s}_m)|\mathcal{S}]\Big)^2\Big|\mathcal{S}\Big].  \nonumberÊ
\end{eqnarray}
{Finally, noting that the term $\mbox{E}_{\widehat{\pi}_m}\big[ \big(h({\bf s}_m)- \mbox{E}_{\widehat{\pi}_m}[h({\bf s}_m)|\mathcal{S}]\big)^2\Big|\mathcal{S}\big]$  is the definition of the variance of the random variable $h({\bf s}_m)$, we obtain}
\begin{eqnarray}
\ell(h)&=&\sum_{m=1}^M\widehat{a}_m^2 \mbox{var}_{\widehat{\pi}_m}[h({\bf s}_m)|\mathcal{S}].
\end{eqnarray}
{Recalling that} $\widehat{a}_m=\sum_{i\in\mathcal{J}_m} \bar{w}_i$  and expressing the variance $\mbox{var}_{\widehat{\pi}_m}[h({\bf s}_m)|\mathcal{S}]$ in terms of weights and samples as shown in Appendix \ref{App2}, we can {also write} the expected loss as
\begin{eqnarray}
\ell(h)&=&\sum_{m=1}^M \left[ \sum_{i\in\mathcal{J}_m} \bar{w}_i  \sum_{i\in\mathcal{J}_m} \bar{w}_i |h({\bf x}_i)|^2-\Big|\sum_{i\in\mathcal{J}_m} \bar{w}_i h({\bf x}_i)\Big|^2 \right], \nonumber \\
\ell(h)&=&\sum_{m=1}^M c_m(h), \label{L2}
\end{eqnarray}
where 
\begin{eqnarray}
\label{C2}
c_m(h)=\sum_{i\in\mathcal{J}_m} \bar{w}_i  \sum_{i\in\mathcal{J}_m} \bar{w}_i |h({\bf x}_i)|^2-\Big|\sum_{i\in\mathcal{J}_m} \bar{w}_i h({\bf x}_i)\Big|^2.
\end{eqnarray}
Note that the expression of $\ell(h)$ {above} is independent from the stochastically-chosen summary particles ${\bf s}_m$. This motivates an
adaptive procedure for building a good partition, as discussed below.
}

\subsection{Compression by kernel density estimation}
 In Eq. \eqref{CMC_PI_Eq}, we can replace the delta functions with kernel functions {$K({\bf x}|{\bf s}_m,{\bm \Sigma}_m)$}, for instance Gaussian kernels $\mathcal{N}({\bf x}|{\bf s}_m,{\bm \Sigma}_m)$, of mean ${\bf s}_m$ and with a $d_X\times d_X$  covariance matrix ${\bm \Sigma}_m$ the $d_X\times d_X$ obtained by an empirical estimation considering the samples in $\mathcal{X}_m$, i.e., 
\begin{eqnarray}
\label{Sigma_m}
{{\bm \Sigma}_m=\sum_{j\in \mathcal{J}_m}  {\bar w}_{m,j} ({\bf x}_j-{\bf s}_m) ({\bf x}_j-{\bf s}_m)^{\top}+\delta {\bf I}},
\end{eqnarray}
where ${\bf s}_m$ is defined in Eq. \eqref{VDet} and $\delta>0$. {Thus, we also have}
\begin{eqnarray}
\label{CMC_PI_KDE_Eq}
{\widetilde \pi}^{(M)}({\bf x})= \sum_{m=1}^M \widehat{a}_m K({\bf x}|{\bf s}_m,{\bm \Sigma}_m),
\end{eqnarray}
where $K(\cdot)$ represents a {so-called} kernel function with location parameter ${\bf s}_m$ and covariance matrix ${\bm \Sigma}_m$.
In a distributed scenario, the $M$ triplets $\{\widehat{a}_m,{\bf s}_m,{\bm \Sigma}_m \}_{m=1}^M$ must be transmitted in the central node.  The transmission of 
$M(\frac{1}{2}d_X^2+\frac{3}{2}d_X+1)$ scalar values are required. 
 { Alternatively, we can use
\begin{eqnarray}
\label{Sigma_m2}
{\bm \Sigma}_m={\bm \Sigma}=\mbox{diag}(\widehat{\sigma}_1^2,....,\widehat{\sigma}_{d_X}^2)+\delta {\bf I},
\end{eqnarray}
where $\widehat{\sigma}_i^2=\mbox{var}_{\widehat{\pi}}[x_{i,n}]$ with $i=1,...,d_X$ and $n=1,...,N$. Hence, only $M(2d_X+1)$ {scalar} values must be transmitted.
}

\subsection{Choice of the partition}
\label{ChoicePartition}
In this section, we discuss some examples of practical choices of the partition, and then a possible adaptive procedure. 
 Given the $N$ samples ${\bf x}_n=[x_{n,1},...,x_{n,d_X}]^{\top}\in \mathcal{D}\subseteq \mathbb{R}^{d_X}$, with $n=1,...,N$. Then, we list three practical choices from the simplest to the more sophisticated strategy:
\begin{itemize}
\item[{\bf P1}]  Random grid, where each component of the elements of the grid are contained  within the intervals $\min\limits_{n=1,...,N} x_{n,i}$  and  $\max\limits_{n=1,...,N} x_{n,i}$, for each $i=1,...,d_X$.
\item[{\bf P2}]  Uniform deterministic grid, where each component of the elements of the grid are contained  within the intervals $\min\limits_{n=1,...,N} x_{n,i}$  and  $\max\limits_{n=1,...,N} x_{n,i}$, for each $i=1,...,d_X$.
\item[{\bf P3}]  Voronoi partition obtained by a clustering algorithm with $M$ clusters (e.g., the well-known $k$-means algorithm). 
\end{itemize}
{\bf Adaptive procedure.} Set $t=0$ and choose an initial partition $\mathcal{P}_0=\{\mathcal{X}_{1}, \mathcal{X}_{2},...,\mathcal{X}_{M_0}\}$ of the domain $\mathcal{D}$, with $M_0=|\mathcal{P}_0|$ disjoint sub-regions, obtained {applying the procedure {\bf P2}}, for instance. Decide also the stopping condition, choosing a maximum number of sub-regions $M_{\texttt{max}}< N$ or  a threshold for the loss, $L$. {Therefore, while $M_t\leq M_{\texttt{max}}$ or $\ell(h)\geq L$ (where $\ell(h)$ is computed as in Eq. \eqref{L1} or  \eqref{L2}),  split the $m^*$-th sub-region, with
\begin{eqnarray}
{m}^*=\arg\max_m    c_m(h).
\end{eqnarray}
Repeat the procedure above{, until the desired stopping condition is reached.} For the stochastic C-MC scheme, this procedure can be extended jointly for several functions $h$. 
}
Recall that we define as {a proper partition rule, {\it any}} partition rule such that when $M=N$, then ${\bf s}_m={\bf x}_n$ and $\widehat{a}_m={\bar w}_n$ (note that $m=n$ in this case), i.e., in the limit case with $M=N$ we consider all the MC samples as summary samples.  
\newline
\newline
{\bf Unweighted C-MC particles.} Let us consider { to have $N$ samples generated by a standard MC or an MCMC algorithm, i.e., we have ${\bar w}_i=\frac{1}{N}$ for $i=1,...,N$}.
 We can choose a partition such that the C-MC weights, $\widehat{a}_m$, are equals. Indeed, if the partition is chosen such that $|\mathcal{J}_m|=\frac{M}{N}$ for all $m$, then $\widehat{a}_m=\frac{1}{M}$. In this case, the partition is related to the empirical quantiles of the target distribution{. In this scenario, we} can interpret the C-MC particles as an approximate quasi-Monte Carlo (QMC) samples. Indeed, as the number of MC samples $N$ grows, the distribution of the nodes ${\bf s}_m$ follows the definition of low-discrepancy \cite{Niederreiter92}. 
Furthermore, since $\widehat{a}_m=\frac{1}{M}$ for all $m$ then, in a distributed scenario, the transmission of summary weights can be avoided: the only information still required is the aggregated weight $W=N$, as we show in the next section. However, we recall that the performance in terms of information loss (see Section \ref{LossSect}) depends on the cost $c_m(h)$ {in} each sub-region.

\section{Application of C-MC and extensions}
\label{DistrSect}

\subsection{Application to distributed inference}
\label{DistrSect2}

\begin{figure}[htbp]
\centering
\includegraphics[width=8cm]{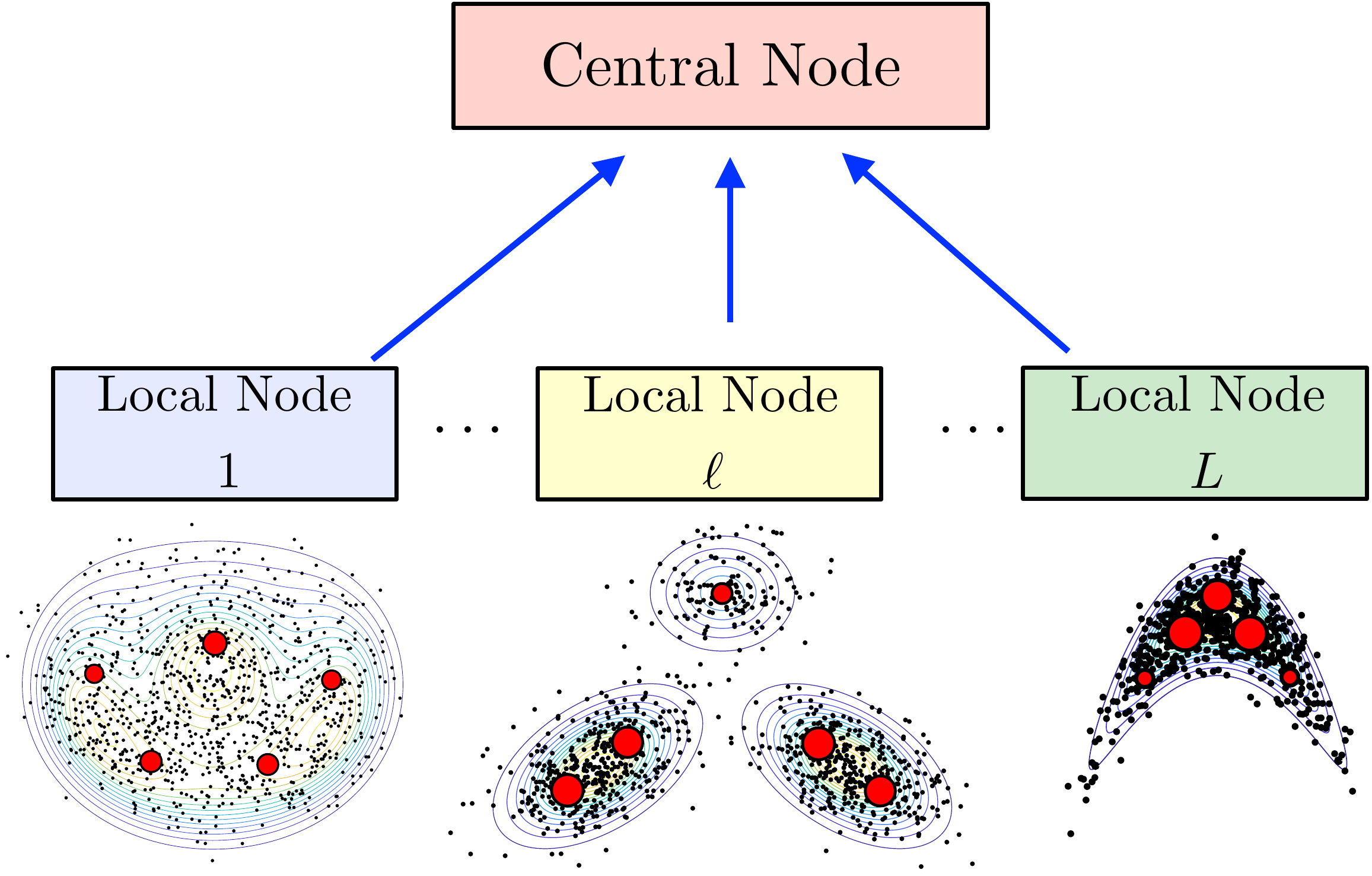}
\caption{Graphical representation of a distributed Bayesian inference framework with $L$ local computational nodes, and a central node. Each local node addresses a posterior density, which is  generally different in each node. If we consider just a parallel framework each node addresses the same posterior. 
}
\label{Fig1teo}
\end{figure}

Distributed algorithms have become a very active topic during the past years favored by fast technological developments (e.g., see \cite{Cetin06}). 
In this section, we consider $L$ independent computational nodes where the Monte Carlo computation is performed in parallel. In the literature, specific techniques have been designed for providing a distributed or diffused inference depending {on whether} a central node is available or not, respectively \cite{Mohammadi16,Farahmand11,Hlinka14}.
Here, we focus on a centralized distributed framework, i.e., we consider a central node where the transmitted local information is properly combined, as represented in Figure \ref{Fig1teo}.  
 We distinguish three different scenarios. In the first one,  from now on {referred to} as {the} parallel framework,  the same dataset ${\bf y}\in \mathbb{R}^{d_Y}$ and the same model is shared by all the local nodes \cite{Bolic05,Read2014,Miguez16}. Thus, all the $L$ nodes address the same inference problem, i.e., they deal with the same posterior density. In the second scenario,  {referred to} as model selection case,  all the nodes have access to the entire dataset ${\bf y}$, but each local node considers a different possible model (different likelihood and/or prior functions), hence they deal with different posteriors \cite{Martino15PF}. The third case is the distributed scenario, {where the} observed data are divided over the $L$ local nodes, ${\bf y}=[{\bf y}_1,...{\bf y}_{L}]^{\top}$. Hence, each node addresses a different sub-posterior density which considers only a subset of the data,  ${\bf y}_{\ell} \in \mathbb{R}^{d_\ell}$ (note that $\sum_{\ell=1}^L d_\ell=d_Y$) \cite{EmbaraMCMC,scott2016bayes}.   
 In these frameworks, a particle compression is often required for reducing the computational {and the} transmission cost. Below, we develop the three frameworks.  
\newline
{\it Parallel framework.}  
We assume the use of $N_\ell$ particles $\{{\bf x}_n^{(\ell)}\}_{n=1}^{N_\ell}$ in each local node. First of all, we consider the transmission of all the particles of the central node, without any compression. In this case, the complete Monte Carlo approximation with $N=\sum_{\ell=1}^L N_\ell$ particles can be expressed as  
\begin{eqnarray}
\label{recall}
{\widehat \pi}_{\texttt{tot}}^{(N)}({\bf x}) 
&=&  \sum_{\ell=1}^L \frac{W_\ell}{\sum_{j=1}^{N_\ell} W_j}\sum_{n=1}^{N_\ell} {\bar \beta}_{n}^{(\ell)} \delta({\bf x}-{\bf x}_n^{(\ell)})  \\
&=& \sum_{\ell=1}^L \bar{\rho}_\ell \ {\widehat \pi}_{\ell}^{(N_\ell)}({\bf x}),
\end{eqnarray}
where $\bar{\rho}_\ell=\frac{W_\ell}{\sum_{j=1}^{N_\ell} W_j}$, and ${\bar \beta}_{n}^{(\ell)}=\frac{1}{N_\ell}$, $W_\ell=N_\ell$ in the case of unweighted samples, or ${\bar \beta}_{n}^{(\ell)}=\frac{1}{N_\ell\widehat{Z}^{(\ell)}} \frac{\pi({\bf x}_n^{(\ell)})}{{q({\bf x}_n^{(\ell)})}}$, $W_\ell=N_\ell\widehat{Z}^{(\ell)}$ in the case of weighted samples. 
Therefore, the complete Monte Carlo approximation ${\widehat \pi}_{\texttt{tot}}^{(N)}({\bf x})$ is a convex combination of the $L$ local particle approximations ${\widehat \pi}_{\ell}^{(N_\ell)}({\bf x})$. If we apply a compression scheme transmitting $M_\ell<N_\ell$ samples, ${\widetilde \pi}_\ell^{(M_\ell)}({\bf x})$ as in Eq. \eqref{CMC_PI_Eq} or \eqref{CMC_PI_KDE_Eq}, then the joint particle approximation in the central node is 
\begin{eqnarray}
\label{compression_centralNode}
{\widetilde \pi}_{\texttt{tot}}^{(M)}({\bf x})= \sum_{\ell=1}^L  \bar{\rho}_\ell \ {\widetilde \pi}_\ell^{(M_\ell)}({\bf x}). 
\end{eqnarray} 
with $M=\sum_{\ell=1}^L M_\ell$.
We aim to have a small loss of information between the particle approximations,  ${\widetilde \pi}_{\texttt{tot}}^{(M)}$ and ${\widehat \pi}_{\texttt{tot}}^{(N)}$.
In \cite{Bolic05,Read2014,Pisland15,GIS18},  the bootstrap strategy described in Section \ref{GoalSect} is applied for the compression. In the numerical experiments, we compare the performance of this strategy with the C-MC approach.
\newline
{{\it Model Selection.}}
The model selection application is an extension of the parallel framework. Indeed, all the nodes process the entire set of data ${\bf y}$, but each local node considers a different possible model $\mathcal{M}_\ell$, hence they address different posterior distributions $\bar{\pi}({\bf x}|{\bf y},\mathcal{M}_\ell)$. In order to tackle this problem, based on the Bayesian Model Averaging (BMA) approach, we need an estimation of the marginal likelihood of each model $\widehat{Z}^{(\ell)}$ (e.g., see \cite{Martino15PF}). For this reason, it is preferable to apply an IS scheme where an estimator of the marginal likelihood is easily provided. { In this scenario, we have again} 
${\widehat \pi}_{\texttt{tot}}^{(N)}({\bf x})  = \sum_{\ell=1}^L \frac{N_\ell\widehat{Z}^{(\ell)}}{\sum_{k=1}^L N_k\widehat{Z}^{(k)}} \ {\widehat \pi}_{\ell}^{(N_\ell)}({\bf x})$  without compression, and ${\widetilde \pi}_{\texttt{tot}}^{(M)}({\bf x})= \sum_{\ell=1}^L  \frac{N_\ell\widehat{Z}^{(\ell)}}{\sum_{k=1}^L N_k\widehat{Z}^{(k)}} \ {\widetilde \pi}_\ell^{(M_\ell)}({\bf x})$, with compression. In this scenario, $ \bar{\rho}_\ell=\frac{N_\ell\widehat{Z}^{(\ell)}}{\sum_{k=1}^L N_k\widehat{Z}^{(k)}}$, for $\ell=1,...,L$, represents an approximation of the posterior probability mass function (pmf) of the model given the data, i.e., $p(\mathcal{M}_\ell|{\bf y})$.
\newline
{\it Distributed framework.}
 {For simplicity,} let us consider $N_\ell=\frac{N}{L}$ and $M_\ell=\frac{M}{L}$, for all $\ell=1,...,L$.  In this case, all the nodes consider the same model as in the parallel scenario, but each local node can process only a portion of the observed data,  ${\bf y}_{\ell} \in \mathbb{R}^{d_\ell}$, with $\sum_{\ell=1}^L d_\ell=d_Y$.  Considering a disjoint subsets of data and a split contribution of the prior as in \cite{EmbaraMCMC}, the complete posterior can be factorized as
\begin{eqnarray}
{\bar \pi}_{\texttt{tot}}({\bf x})\propto \prod_{\ell=1}^L {\bar \pi}_{\ell}({\bf x}).
\end{eqnarray}
In different works  \cite{EmbaraMCMC,scott2016bayes}, local approximations of the sub-posteriors are provided and transmitted to the central node, obtaining 
\begin{eqnarray}
\label{DistrScenarioPI}
{\widehat \pi}_{\texttt{tot}}^{(N)}({\bf x})\propto \prod_{\ell=1}^L {\widehat \pi}_{\ell}^{(N_\ell)}({\bf x}).
\end{eqnarray} 
The simplest approach considers Gaussian local approximations \cite{EmbaraMCMC,scott2016bayes}.
A more sophisticated approach proposed in \cite[Section 3.2]{EmbaraMCMC} considers a mixture of Gaussian pdfs as KDE local approximation using all the $N_\ell=\frac{N}{L}$ samples in each node, i.e.,
\begin{eqnarray}
\label{EmbarrassingEQ}
{\widehat \pi}_\ell^{(N_\ell)}({\bf x})= \sum_{n=1}^{N_\ell} {\bar \beta}_{n}^{(\ell)} \mathcal{N}({\bf x}|{\bf x}_{n}^{(\ell)},\delta {\bf I}),
\end{eqnarray}
with $\delta >0$ and ${\bf I}$ is a $d_X \times d_X$ identity matrix. It is easy to see that ${\widehat \pi}_{\texttt{tot}}^{(N)}({\bf x})$ in Eq. \eqref{DistrScenarioPI} can be expressed as a mixture of $N_\ell^L$  Gaussian components \cite{EmbaraMCMC,NIPS2003_2435}. It is possible to draw from this mixture {of densities}, but clearly the cost depends of the number of $N_\ell^L$ components \cite{NIPS2003_2435}. Therefore, here the advantage of using a compressed local mixture, ${\widetilde \pi}^{(M_\ell)}({\bf x})= \sum_{m=1}^{M_\ell} \widehat{a}_m \mathcal{N}({\bf x}|{\bf s}_m,{\bm \Sigma}_m)$ with $M_\ell<N_\ell$, is even more apparent than in the parallel scenarios described above. Indeed, using C-MC, we obtain ${\widehat \pi}_{\texttt{tot}}^{(M)}({\bf x})\propto \prod_{\ell=1}^L {\widetilde \pi}_{\ell}^{(M_\ell)}({\bf x})$, that can be expressed as a mixture of $M_\ell^L$  Gaussian pdfs  \cite{EmbaraMCMC,NIPS2003_2435}.

{
\subsection{Application to particle filtering}

In this section, we show how C-MC can be employed for a performance improvement or a decrease of the computational cost {within particle filtering (PF) algorithms}. Let us consider the following state-space model
\begin{equation}
\left\{
\begin{array}{l}
{\bf x}_{t}\sim p({\bf x}_{t}|{\bf x}_{t-1}) \\
  {\bf y}_t\sim p({\bf y}_{t}|{\bf x}_{t})
\end{array}
\right.,
\qquad t=1,\ldots, T,
\end{equation}
described by the propagation kernel,  $p({\bf x}_{t}|{\bf x}_{t-1})$, and the likelihood function $p({\bf y}_{t}|{\bf x}_{t})$.
Below, we provide two novel schemes based on C-MC.
\newline
{\bf Improved Gaussian particle filter (I-GPF).} The Gaussian particle filter (GPF) is a well-known benchmark PF algorithm proposed in \cite{Kotecha03a}.
The GPF outperforms of conventional Gaussian filters (like the Extended Kalman filter and its variants) in many scenarios and presents lower complexity than standard particle filters.  The resampling steps In the GPF are replaced by a sampling step from an adapted Gaussian density. Table \ref{IGPFtable} describes {the} novel scheme based on C-MC{,} where the pdf in Eq. \eqref{CMC_PI_KDE_Eq} plays the role of the Gaussian density in the standard GPF. Note that, with $M=1$, we recover the standard GPF whereas, with $M=N$, the I-GPF is equivalent to the well-known regularized particle filter \cite{SMC01}. {Moreover,} resampling from ${\widehat \pi}^{(N)}$ is more costly than resampling from ${\widetilde \pi}^{(M)}$ if $M<N$.  {Related} ideas can {be found} in the literature \cite{Kotecha03b,Li_2012}.  The performance of GPF and I-GPF are compared in Section \ref{GPFsect}.

\begin{table}[!h]
{
\caption{\normalsize Improved Gaussian particle filter (I-GPF)}
\label{IGPFtable}
\vspace{-0.3cm}
\begin{tabular}{|p{0.95\columnwidth}|}
   \hline
{\bf Initialization:} Choose $N$, $M$ and $\bar{\bf x}_{0}^{(i)}$, with $i=1,...,N$.\\
{\bf For $t=1,...,T:$}
\begin{enumerate}
\item Draw ${\bf x}_t^{(i)} \sim p({\bf x}_t |\bar{\bf x}_{t-1}^{(i)})$, with $i=1,...,N$.
\item Compute the $M$ weights
\begin{equation}
w_n= p({\bf y}_t|{\bf x}_t^{(i)}), \quad i=1,...,N.
\end{equation}
\item Apply a C-MC scheme for obtaining ${\widetilde \pi}^{(M)}({\bf x})$ in Eq. \eqref{CMC_PI_KDE_Eq}.
\item Draw $\bar{\bf x}_t^{(n)}\sim {\widetilde \pi}^{(M)}({\bf x})$  with $n=1,...,N$.
\end{enumerate}  \\
\hline 
\end{tabular}
}
\end{table}

\noindent
{\bf Compressed  particle filter (C-PF).}  If the compression is applied before the evaluation of the likelihood function $p({\bf y}_t|{\bf x}_t)$, we have an additional reduction of the computational cost. Indeed, in this case, we need to evaluate the { likelihood,} only $M<N$ times at the summary particles ${\bf s}_m$.  
This is particularly convenient if the evaluation of the likelihood is costly (due to the number of data, {or} a complex measurement model). The C-PF is given in Table \ref{CPFtable}. As in I-GPF, the resampling step is performed over $M$ weighted samples instead of $N$. Thus, {C-PF is cheaper} and faster than a standard particle filter. Note that the C-MC weights ${\widehat a}_m$ are included in particle weights in Eq. \eqref{AquiAm}. The weighted points  $\{{\bf s}_m,{\widehat a}_m\}_{m=1}^M$ play a similar role than the sigma points in the Unscented Kalman filter (UKF)  \cite{Julier04,Sarkka13bo}. 
\newline
Other possible applications of C-MC are within the so-called parallel partitioned particle filters and multiple particle filters, as {an alternative} to the use of first moment estimators (or sigma points) for approximating marginal posterior distributions \cite{MPF}. Similar ideas has been also applied within particle smoothing techniques \cite{Klaas06}.  

\begin{table}[!h]
{
\caption{\normalsize Compressed Particle Filter (C-PF)}
\label{CPFtable}
\vspace{-0.3cm}
\begin{tabular}{|p{0.95\columnwidth}|}
   \hline
{\bf Initialization:} Choose $N$, $M$ and $\bar{\bf x}_{0}^{(i)}$, with $i=1,...,N$.\\
{\bf For $t=1,...,T:$}
\begin{enumerate}
\item Draw ${\bf x}_t^{(i)} \sim p({\bf x}_t |\bar{\bf x}_{t-1}^{(i)})$, with $i=1,...,N$.
\item Apply a C-MC scheme  obtaining  $\{{\bf s}_m,{\widehat a}_m\}_{m=1}^M$.
\item Compute the $M$ weights
\begin{equation}
\label{AquiAm}
w_m={\widehat a}_m p({\bf y}_t|{\bf s}_m), \quad m=1,...,M.
\end{equation}
and normalized them $\bar{w}_m=\frac{w_m}{\sum_{k=1}^M w_k}$.
\item Obtain $\{\bar{\bf x}_t^{(n)}\}_{n=1}^N$, by resampling $N$ times within $\{{\bf s}_m\}_{m=1}^M$ according to $\bar{w}_m$, with $m=1,...,M$.
\end{enumerate} \\ 
\hline 
\end{tabular}
}
\end{table}

\subsection{Application to adaptive importance sampling}

 In the so-called  layered adaptive importance sampling (LAIS) algorithm \cite{LAIS17} and similar methods \cite{Ingmar}, an MCMC algorithm is used for obtaining a set of mean parameters $\{{\bm \mu}_1,...,{\bm \mu}_T\}$. Then, one sample ${\bf x}_t$  is drawn from a proposal density with mean ${\bm \mu}_t$, i.e., ${\bf x}_t \sim q({\bf x}_t|{\bm \mu}_t,{\bf C})$ where ${\bf C}$ is a covariance matrix and $t=1,....,T$. One possible choice of the weights is
  \begin{equation}
\label{Weq2}
w_t=\frac{\pi({\bf x}_t)}{\frac{1}{T} \sum_{k=1}^T q({\bf x}_t|{\bm \mu}_k,{\bf C})},
\end{equation}
where a temporal mixture is used in the denominator \cite{LAIS17,Ingmar}.  With this choice, very good performance can be obtained,  but the computational cost of evaluating the weight denominator increases with $T^2$ {\cite{LAIS17}}. 
 If $T$ is large, the evaluation of the weights in Eq. \eqref{Weq2} can be {costly}. 
Hence, the C-MC scheme can be applied to the set $\{{\bm \mu}_1,...,{\bm \mu}_T\}$ are shown in Table \ref{CLAIStable}. More generally, C-MC can be also applied within adaptive MC schemes {to} obtain a good construction of the adaptive proposal density \cite{cappe04,pmc-cappe08,7974876}.


\begin{table}[!h]
{
\caption{\normalsize Compressed LAIS (CLAIS)}
\label{CLAIStable}
\vspace{-0.3cm}
\begin{tabular}{|p{0.95\columnwidth}|}
   \hline
\begin{enumerate}
\item  Generate a chain ${\bm \mu}_1,...,{\bm \mu}_T$ using an MCMC technique (with target $\pi$ or a tempered version).
\item  Draw $T$ samples from ${\bf x}_t \sim q({\bf x}_t|{\bm \mu}_t,{\bf C})$, with $t=1,...,T$, and where ${\bf C}$ is a covariance matrix.
\item  Considering the samples $\{{\bf x}_{t}\}_{t=1}^T$, obtain ${\widetilde \pi}^{(M)}({\bf x})$ in Eq. \eqref{CMC_PI_KDE_Eq} by C-MC,  with $M < T$.
\item  To each ${\bf x}_t$, assign the weight
\begin{equation}
\label{Weq}
w_t=\frac{\pi({\bf x}_t)}{{\widetilde \pi}^{(M)}({\bf x}_t)}.
\end{equation}
\end{enumerate} \\ 
\hline 
\end{tabular}
}
\end{table}

}

\vspace{-0.9cm}

{
\subsection{Extensions: Least Squares CMC  (LS-CMC)}
If we relax the assumption {that the weights} $\widehat{a}_m$ must be non-negative, we can obtain better performance in terms of loss in compression.
Indeed, given the summary particles $\{{\bf s}_m\}_{m=1}^M$ considering a family of $R+1$ functions, i.e., $\mathcal{H}=\{h_0({\bf x})=1,h_1({\bf x}),..., h_R({\bf x})\}$, we can write the {following} linear system, 
\begin{gather}\label{SystemLSweights}
\left\{
\begin{split}
 &\sum_{m=1}^M  \widehat{a}_m=1,Ê\\
 &\sum_{m=1}^M  \widehat{a}_m h_1({\bf s}_m)=\widehat{I}^{(N)}(h_1), \\
 &\quad \vdots \\
  &\sum_{m=1}^M  \widehat{a}_m h_R({\bf s}_m)=\widehat{I}^{(N)}(h_R).
\end{split}
\right. 
\end{gather}
{ with $M$ unknowns and $R+1$ equations.}
If $M\leq  R+1$ the system is overdetermined, and it has in general no solution. However, we can {still} find a {Least Squares (LS)} solution  {for this problem. Indeed,}  the system {in Eq. \eqref{SystemLSweights}} can be rewritten as
$$
{\bf H} {\bf \widehat{a}}\approx {\bf v},
$$
where $ {\bf H}$ is a $(R+1)\times M$ matrix with entries ${\bf H}_{ij}=h_i({\bf s}_j)$, ${\bf a}=[\widehat{a}_1,...,\widehat{a}_M]^{\top}$ is the vector of the unknowns, and ${\bf v}=[1,\widehat{I}^{(N)}(h_1),...,\widehat{I}^{(N)}(h_R)]^{\top}$. {The well-known LS} solution is {then} given by 
\begin{equation}
{\bf \widehat{a}}=({\bf H}^{\top}{\bf H})^{-1} {\bf H}^{\top} {\bf v}.
\end{equation}
Note that the weights {in the vector} ${\bf \widehat{a}}=[\widehat{a}_1,...,\widehat{a}_M]^{\top}$ {could} be also negative. For this reason, the range of application of LS-CMC is reduced but, for instance, LS-CMC can be still applied to the pure parallel framework{,} described in Section \ref{DistrSect2}.

}




\section{Numerical experiments}
\label{Simu}

In the section, we test the proposed C-MC techniques in six different numerical examples and compare their performance with the corresponding benchmark methods. In the first experiment, we apply the compression techniques to two sets of  Monte Carlo samples. In the second experiment, we consider a localization problem in a wireless sensor network and the use of $L$ local parallel processors. We test the performance of the Compressed LAIS (CLAIS) scheme for {performing an inference} in {an exoplanetary} model, in the third example. The last three experiments consider the use of particle filtering. In Section \ref{CPFsimu}, we test the proposed C-PF obtaining very promising results. Finally, in Sections \ref{GPFsect} and \ref{DistribuidoPF}, we consider two different object tracking problems with different measurements and propagation models. Moreover, in Section \ref{DistribuidoPF} a centralized distributed inference problem is considered.

\subsection{First numerical analysis}

\begin{figure}[htbp]
\centering
\subfigure[Gamma target pdf]{\includegraphics[width=8cm]{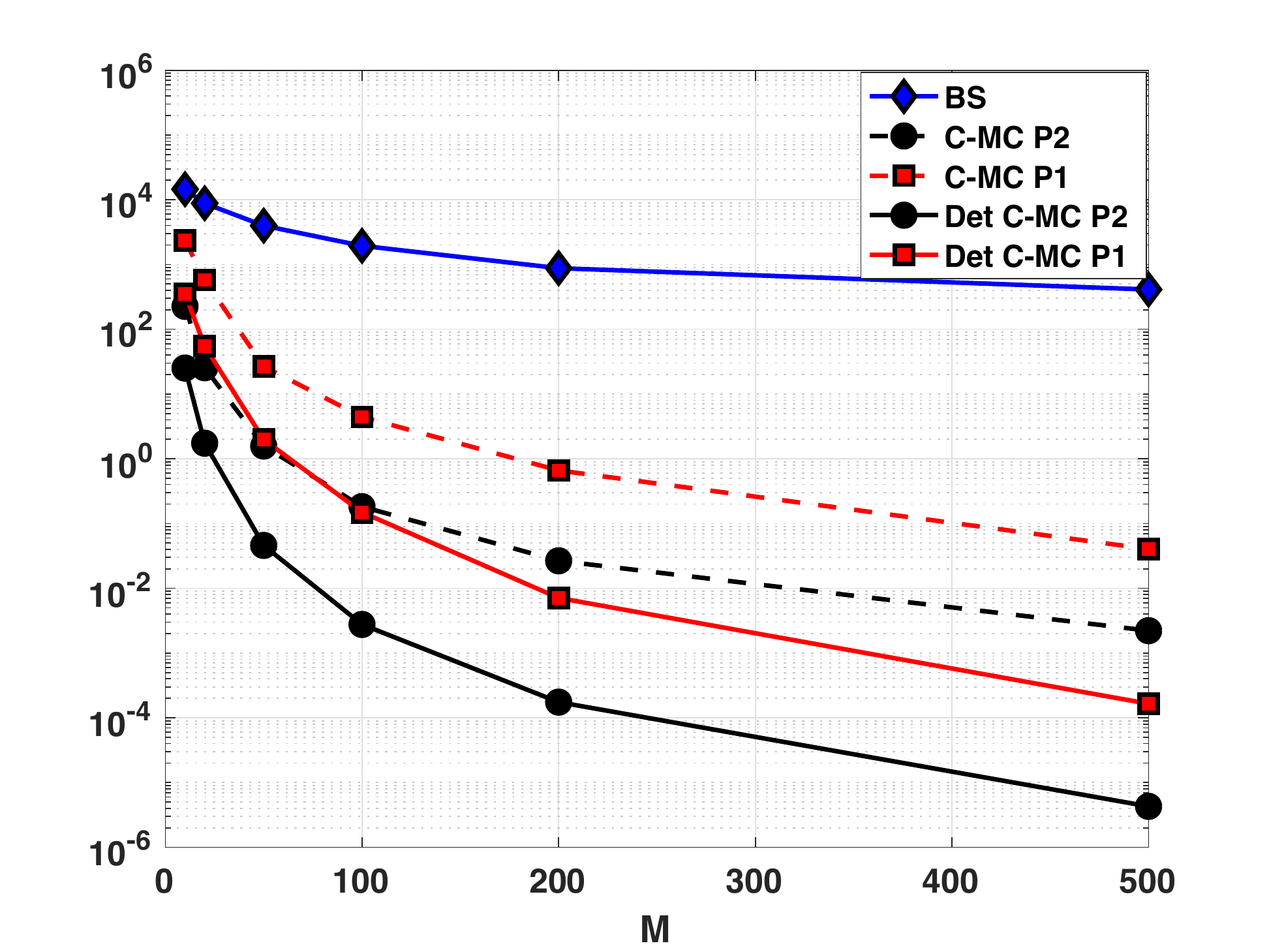}}
\subfigure[Mixture target pdf]{\includegraphics[width=8cm]{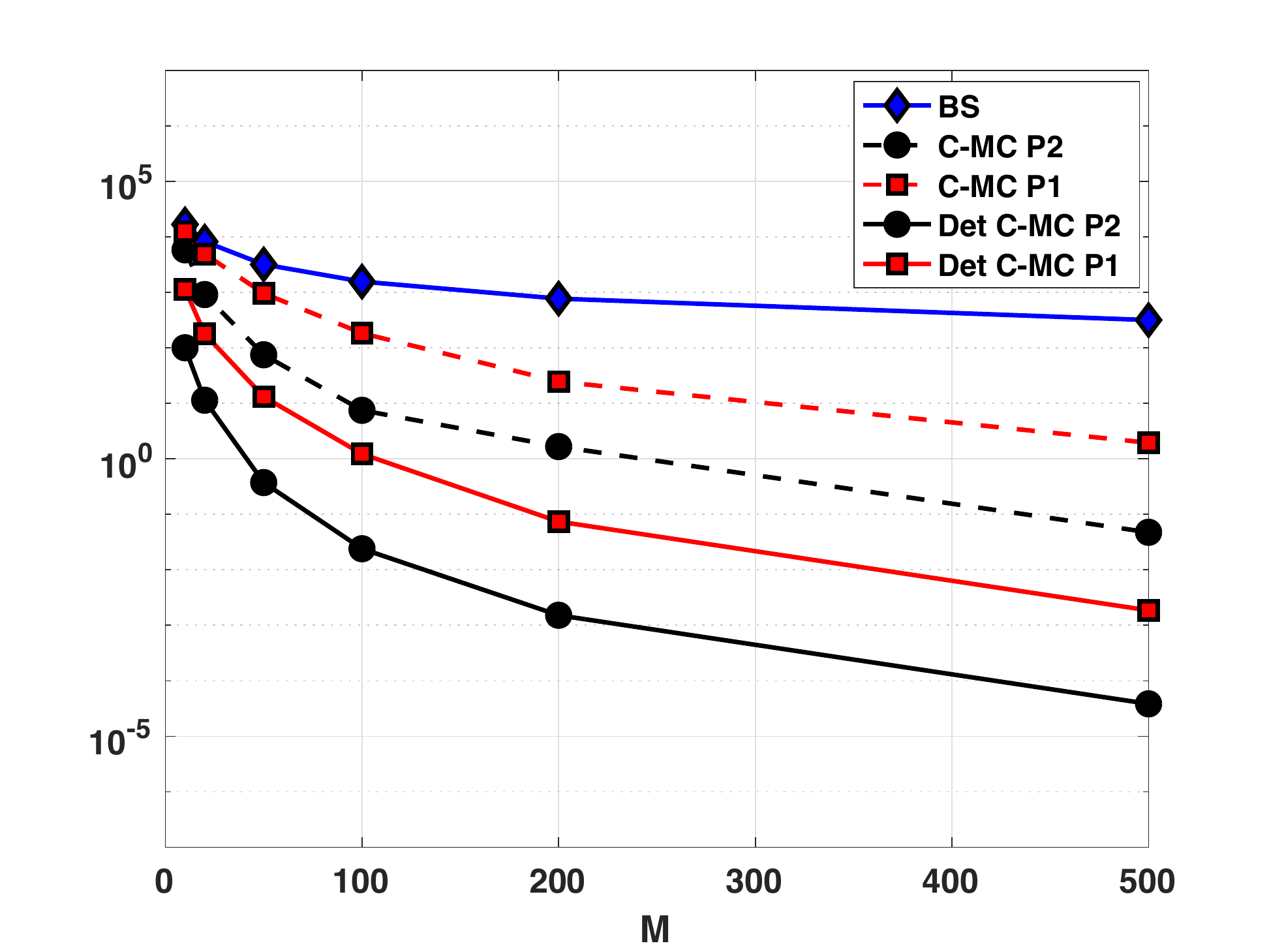}}
\caption{The loss $\mathcal{L}_5$ as function of $M$. The results obtained by the bootstrap strategy  \cite{Bolic05,Read2014,GIS18} in Section \ref{GoalSect} is depicted with a solid line and rhombuses. The results of C-MC with a random partition (P1){, and with} a grid partition (P2) are shown by squares and circles, respectively. The results obtained with the deterministic choice of ${\bf s}_m$ in Eq. \eqref{VDet} are shown with solid lines (squares and circles), whereas the {results corresponding to the random choice} of ${\bf s}_m$ are provided with dashed lines (squares and circles). }
\label{FigEX1}
\end{figure}

Let {start, for simplicity, with a scalar scenario, i.e.,} $x\in \mathbb{R}$. {Furthermore,} we consider two possible target densities: the first one is a Gamma pdf
\begin{equation}
\bar{\pi}(x)Ê\propto x^{\alpha-1}\exp\left(-\frac{x}{\kappa}\right),
\end{equation}
with $\alpha=4$ and $\kappa=0.5$, and the second one is a mixture of two Gaussians,
\begin{equation}
\bar{\pi}(x)Ê= \frac{1}{2}\mathcal{N}(x|-2,1)+\frac{1}{2}\mathcal{N}(x|4,0.25).
\end{equation}
{\it Experiments.} We generate $N=10^5$ Monte Carlo samples from both and compare the bootstrap strategy (BS) with different C-MC schemes. More specifically, we consider two kind of partition procedures:  random (P1) and uniform (P2) described in Section \ref{ChoicePartition}.  {Furthermore,} we compare the stochastic and the deterministic choices of the summary particles ${\bf s}_m${,} described in Section \ref{SectCMC}. Therefore, for the deterministic C-MC {schemes,} we  {consider} the use {of  ${\bf s}_m$ in Eq. \eqref{VDet}}. We repeat the experiment $500$ independent runs and  average the results.
At each run, we compute the loss $\mathcal{L}_5$ with $\xi_r^2=1$, for $r=1,...,5$ (i.e., the loss in the first $5$ moments) provided by the different techniques. Figure \ref{FigEX1} depicts the averaged $\mathcal{L}_5$ as function of the number $M$ of summary particles. Figure \ref{FigEX1}-{\bf (a)} refers to the Gamma target pdf, whereas Figure \ref{FigEX1}-{\bf (b)} corresponds to the Gaussian mixture pdf. The results of the BS method are displayed with triangles. The stochastic C-MC schemes are shown with dashed lines, whereas the deterministic C-MC schemes with solid lines. 
\newline
{\it Discussion.} In all cases, C-MC outperforms BS and the deterministic C-MC schemes provide {the best} results. {As expected,} the partition P2 (depicted with circles) outperforms P1 (shown with squares).  Note that P1 represents the simplest and perhaps the worst possible construction of the partition. However, it is important to remark that the C-MC schemes, even with P1, outperform the BS method.\footnote{{The code} of this first example is provided at \url{http://www.lucamartino.altervista.org/CMC_CODE_pub_EX1.zip}.}

\subsection{Localization in a sensor network with Parallel AIS schemes}
\label{PAIS}

\begin{figure*}[htbp]
\centering
\centerline{
\subfigure[]{\includegraphics[width=4.6cm]{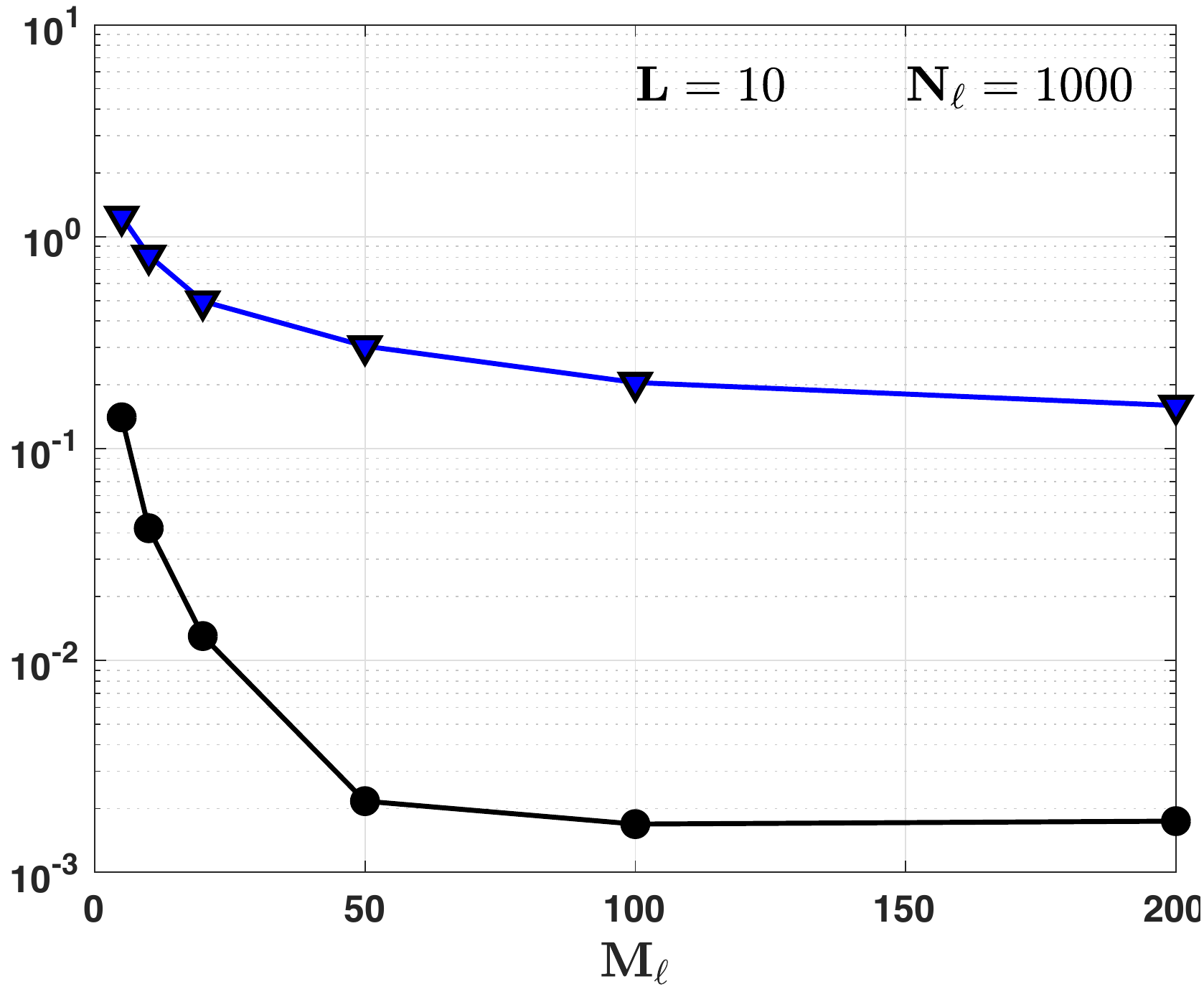}}
\subfigure[]{\includegraphics[width=4.6cm]{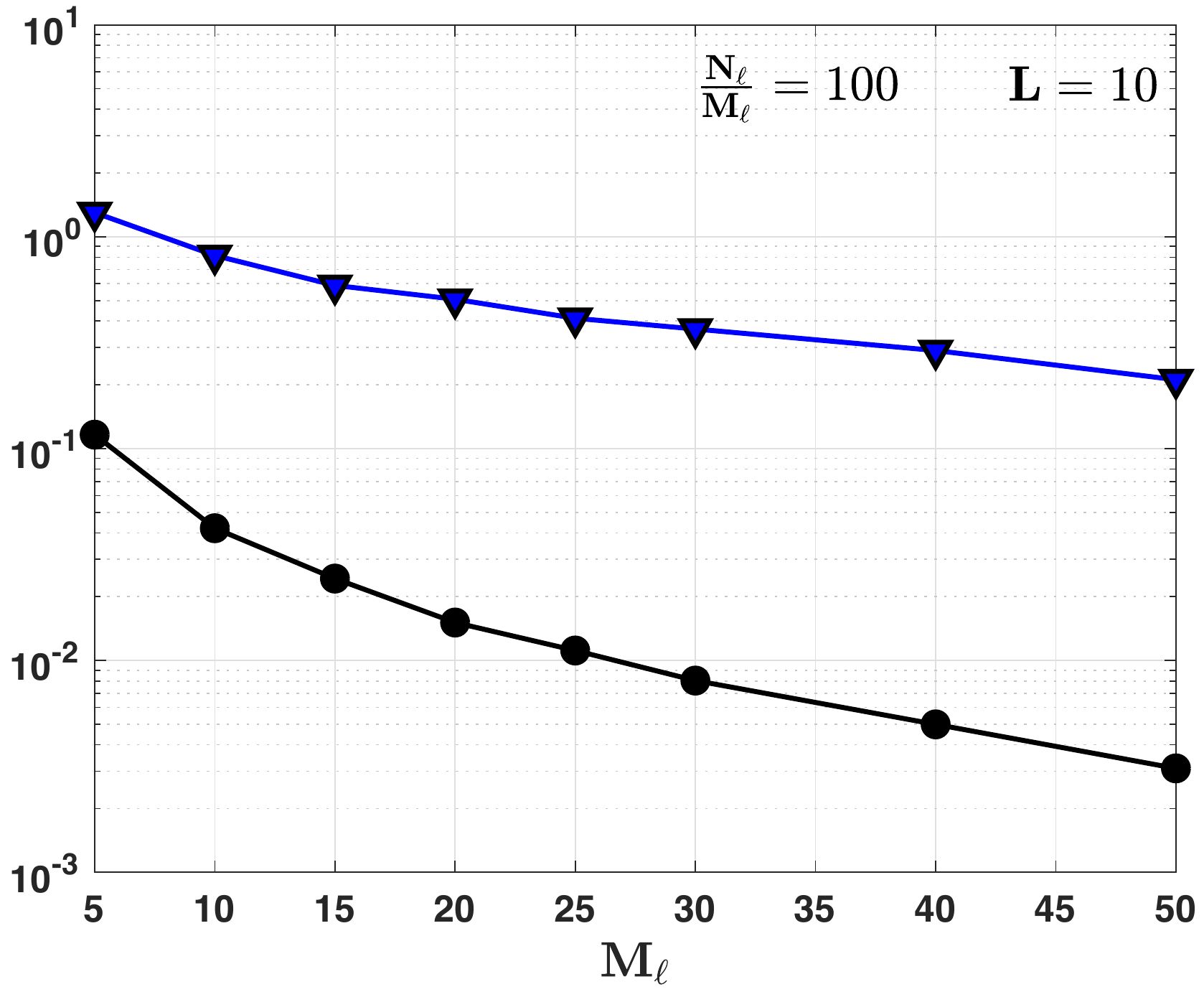}}
\subfigure[]{\includegraphics[width=4.7cm]{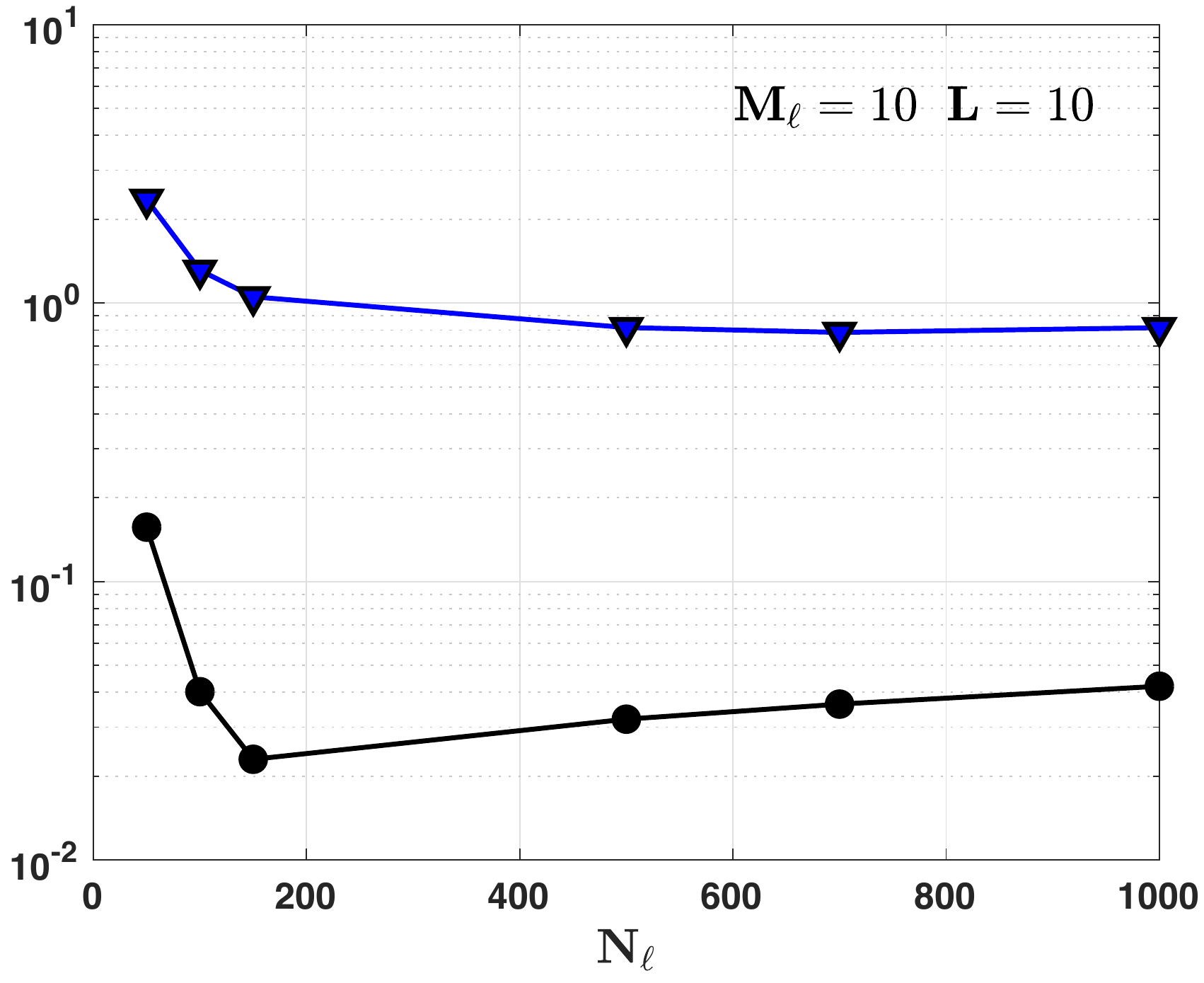}}
\subfigure[]{\includegraphics[width=4.6cm]{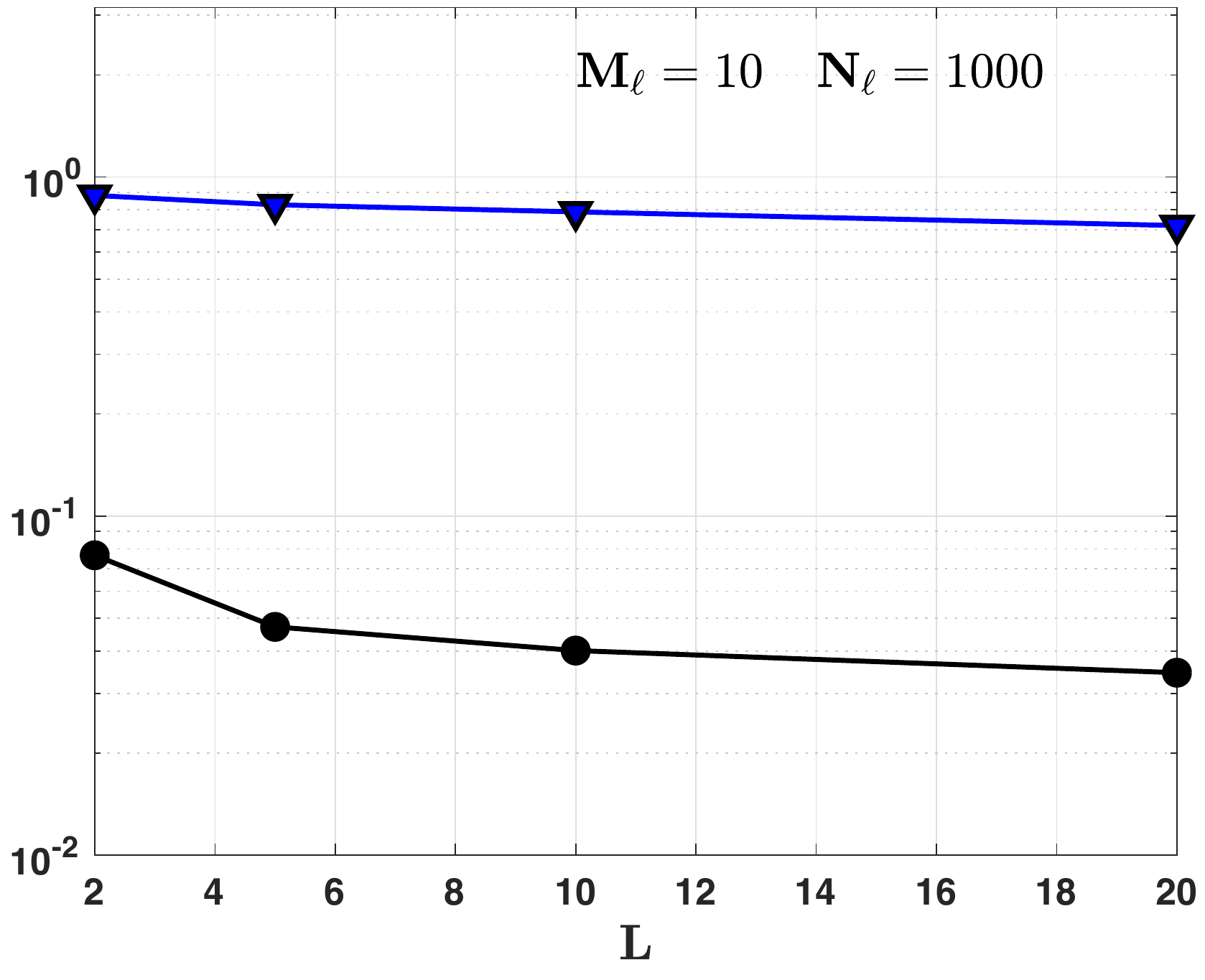}}
}
\caption{{\bf (a)}-{\bf (b)}-{\bf (c)}-{{\bf (d)}} Results in terms of information loss for the localization problem in wireless sensor network: C-MC is shown with circles and the bootstrap strategy with triangles. 
}
\label{FigEX2}
\end{figure*}

In this section, we test the C-MC technique considering the problem of positioning a target in $\mathbb{R}^{2}$ using a range measurements in a wireless sensor network \cite{Ihler05}. 
Specifically, the target position is modeled as a random vector $\textbf{X}=[X_1,X_2]^{\top}$, hence the actual position of the target is a specific realization ${\bf X}={\bf x}$. 
The data (range measurements) are obtained from $3$ sensors located at $\textbf{h}_1=[3, -8]^{\top}$, $\textbf{h}_2=[10,0]^{\top}$, $\textbf{h}_3=[0,10]^{\top}$, as shown in Figure \ref{FigEX2}-{\bf (d)}. The likelihood function is induced by the following observation model,
\begin{gather}
\label{IStemaejemplo}
\begin{split}
Y_{j}=20\log\left(||{\bf x}-{\bf h}_j ||\right)+B_{j}, \quad j=1,2, 3, \\
\end{split}   
\end{gather}   
where $B_{j} \sim \mathcal{N}(b_j;0,\lambda_j^2)$. 
We consider the true position of the target as ${\bf x}^*=[x_1^*=2.5,x_2^*=2.5]^{\top}$ and set $\lambda_j=6$. 
Then, we generate one measurement $y_j$ from each sensor according to the model in Eq. \eqref{IStemaejemplo}, obtaining the vector ${\bf y}=[y_1,y_2, y_3]$.  Assuming a uniform prior in the rectangle $\mathcal{R}_z=[-30, 30]^2$, then  the posterior density is 
 {\small
\begin{eqnarray}
\label{PostEx2}
{\bar \pi}({\bf x})\propto\left[\prod_{j=1}^{3} \exp\left(-\frac{1}{2\lambda_j^2}(y_{j}-20\log\left(||{\bf z}-{\bf h}_j ||\right)^2\right) \right]\mathbb{I}_{\mathcal{R}_z}({\bf x}), 
\end{eqnarray}
}
where $\mathbb{I}_{\mathcal{R}_z}({\bf x})$ is an indicator function that is $1$ if ${\bf x}\in\mathcal{R}_z$, otherwise is $0$. 
\newline
{\it Parallel setup.} We assume $L$ local computational nodes. At each one, we run an adaptive importance sampler, specifically a standard Population Monte Carlo (PMC) scheme \cite{cappe04}. Each PMC delivers $N_\ell$ weighted samples as an approximation of the posterior of Eq. \eqref{PostEx2}, after a certain number of iterations \cite{7974876}. Therefore, we have $\widehat{\pi}_\ell^{(N_\ell)}$ local approximations of $N_\ell$ particles. In this setting, we have a clear improvement in term of computational times{, since the} $L$ different PMC algorithms are run in parallel. When all the samples are transmitted to the central node, we obtain a complete particle approximation $\widehat{\pi}_{\texttt{tot}}^{(N)}$ as in Eq. \eqref{recall} with $N=\sum_{\ell=1}^L N_\ell$ (we set $N_\ell=\frac{N}{L}$). However, in general due to the transmission cost, a particle compression is applied. 
 In this case, we have $L$ local approximations  $\widetilde{\pi}_\ell^{(M_\ell)}$, and the central node performs the fusion obtaining $\widetilde{\pi}_{\texttt{tot}}^{(M)}$ as in Eq. \eqref{compression_centralNode} with $M=\sum_{\ell=1}^L M_\ell$ (we set $M_\ell=\frac{M}{L}$).
We measure the quality of the approximation $\widetilde{\pi}_{\texttt{tot}}^{(M)}$ computing the loss (i.e., mean square error) in the estimation of the mean vector, the covariance matrix, skewness, and kurtosis vectors  (i.e., overall $9$ scalar values) with respect to $\widehat{\pi}_{\texttt{tot}}^{(N)}$. We compare the bootstrap strategy (BS) {in} \cite{Bolic05,Read2014,Pisland15,GIS18} and C-MC. For building the partition for C-MC,  {we perform a k-means clustering with $M_\ell$ clusters in each local node. The clustering is applied after resampling $N_\ell$ times within the weighted particles given by PMC.} 
Thus, the partition is given by the  $M_\ell$ Voronoi regions. Then, we consider again the weighted samples produced by the PMC and build the summary weights $\widehat{a}_m$ and summary samples ${\bf s}_m$ for each Voronoi region. We average the results over 200 independent runs.
\newline
{\it Experiments.}  The losses of BS (triangles) and C-MC (circles) for different values of  $M_\ell$ and $N_\ell$ (witht $L=10$) are depicted in Figures \ref{FigEX2} {\bf (a)}-{\bf (b)}-{\bf (c)}. More specifically, in Figure \ref{FigEX2}-{\bf (a)} we set $N_\ell=1000$ and vary $M_\ell$.
In Figure \ref{FigEX2}-{\bf (b)}, we vary $M$ keeping fixed the compression rate $\eta=\frac{N_\ell}{M_\ell}=100$, i.e., when $M_\ell$ grows also $N_\ell$ is increased. In Figure \ref{FigEX2}-{\bf (c)}, we set $M_\ell=10$, and vary $N_\ell$.  { Finally, in Figure \ref{FigEX2}-{\bf (d)} we set $M_\ell=10$, $N_\ell=1000$ and vary $L$.}
\newline
{\it Discussion.} First of all, we can observe that C-MC always outperforms BS providing the small loss in any scenario. The increase of $M_\ell$ has always a positive impact as shown in Figures \ref{FigEX2}-{\bf (a)}-{\bf (b)}. In Figure \ref{FigEX2}-{\bf (c)}, the compression rate $\eta=\frac{N_\ell}{M_\ell}$ is increasing since $M_\ell$ is fixed and $N_\ell$ grows, so that we expect that the performance should become worse as $N_\ell$ grows. However, in a first moment, the increase of $N_\ell$ {helps both} schemes, C-MC and BS, since a better partition can {be built} with a greater $N_\ell$ in C-MC by clustering, and the resampling steps used in bootstrap improves its performance with a greater $N_\ell$ in BS. Moreover, in this scenario, the increase of $N_\ell$ seems to have  {a more positive} impact {on} the BS technique. 
However, Figure \ref{FigEX2}-{\bf (b)} shows that, if the compression rate $\eta=\frac{N_\ell}{M_\ell}$ is maintained fixed, then C-MC obtains a better improvement. In Figure \ref{FigEX2}-{\bf (d)}, we can see that the performance improves when $L$ grows. 


{
\subsection{Inference in a exoplanetary model} 
In this section, we consider the application of the Compressed LAIS (CLAIS) scheme described in Table \ref{CLAIStable} to make inference in {an exoplanetary} system.
Let us consider the following simplified observation model of a Keplerian orbit and the radial velocity of the host star,
\begin{equation}
y_j= V + \sum\limits_{i = 1}^{N_P} K_i \left[ \cos \left(\frac{2\pi}{P_i} t_j+ \omega_i \right) + e_i \cos \left( \omega_i \right) \right]+ \xi_j,
\label{eq:rv}
\end{equation}
where $y_j$ is the $j$-th observation, $t_j$ is a known time instants, $V$ is the mean radial velocity, $N_p$ is the  number of planets in the system, and $K_i$ is an  amplitude, $P_i$ is the  period, $\omega_i$
is  longitude of periastron, $e_i$ the eccentricity of the orbit and $ \xi_j\sim \mathcal{N}(0,1)$ \cite{Balan09}. 
We consider that all the parameters $V$, $K_i$, $P_i$, $e_i$, $\omega_i$ are unknown for $i=1,...,N_P$  and also  the number of planets $N_P$ is unknown. 
Note that the dimension of the inference space depends on $N_P$: if there is no planet in the system ${\bf x}=V$ then $d_X=1$, with $N_P=1$ we have ${\bf x}=[V,K_1,P_1,e_1, \omega_1]^{\top}$ then $d_X=5$, with $N_P=2$ we have ${\bf x}=[V,K_1,P_1,e_1, \omega_1,K_2,P_2,e_2, \omega_2]^{\top}$ hence $d_X=9$, i.e., generally we have $d_X=1+5N_P$.

{Let consider} $50$ data stacked in a vector ${\bf y}${, generated} from the model in Eq. \eqref{eq:rv}{. Our goal} is to make inference regarding the number of $N_P$ and the corresponding parameters, {with $0<N_P\leq 3$}. We consider uniform priors $\mathcal{U}([a,b])$ over the parameters ($a=-20$, $b=20$ for $V$, $a=0$, $b=365$ for $P_i$, $a=-\pi$, $b=\pi$ for $\omega_i$, $a=0$, $b=1$  for $e_i$) and a uniform discrete prior $p_i=1/4$ over the number of planets, $N_P$. {We fix $N_P$,} and apply CLAIS with a random walk Metropolis chain \cite{Robert04}, of length $T=20^5$ and set $M=10$ (see Table \ref{CLAIStable}). The partition is built by the approach P2 given in Section \ref{ChoicePartition}. With CLAIS we can easily estimate the marginal likelihood $\widehat{Z}^{(i)}$ with $i=0,...,3${, using the corresponding IS estimator.} Then, the marginal posterior of $N_P$ is approximated by
\begin{eqnarray}
p(N_P=k |{\bf y})\approx \frac{\widehat{Z}^{(k)}}{\sum_{i=0}^3 \widehat{Z}^{(i)}}, 
\end{eqnarray}
with $k=0,...,3$. We make two experiments. First, we set  $N_P=1$ and then $N_P=3$ planets {and generate the corresponding data ${\bf y}$.} Note that for computing $p(N_P=k |{\bf y})$ we need to integrate out the rest of parameters. The probabilities $p(N_P=k |{\bf y})$ obtained in the two experiments are given in Figure \ref{FigEX_CLAIS}. Note that the task of providing a good estimation of $\widehat{Z}^{(i)}$ depends on the ability of the sampling method of exploring properly the state space. For this reason, the need of increasing the length $T$ of the MCMC chain raises as the dimension $d_X$ grows. In Figure \ref{FigEX_CLAIS}, we can observe that CLAIS is able to recover the number of planets in each experiments{. The results are averaged over 100 independent runs.}

   \begin{figure}[htbp]
\centering
\includegraphics[width=4cm]{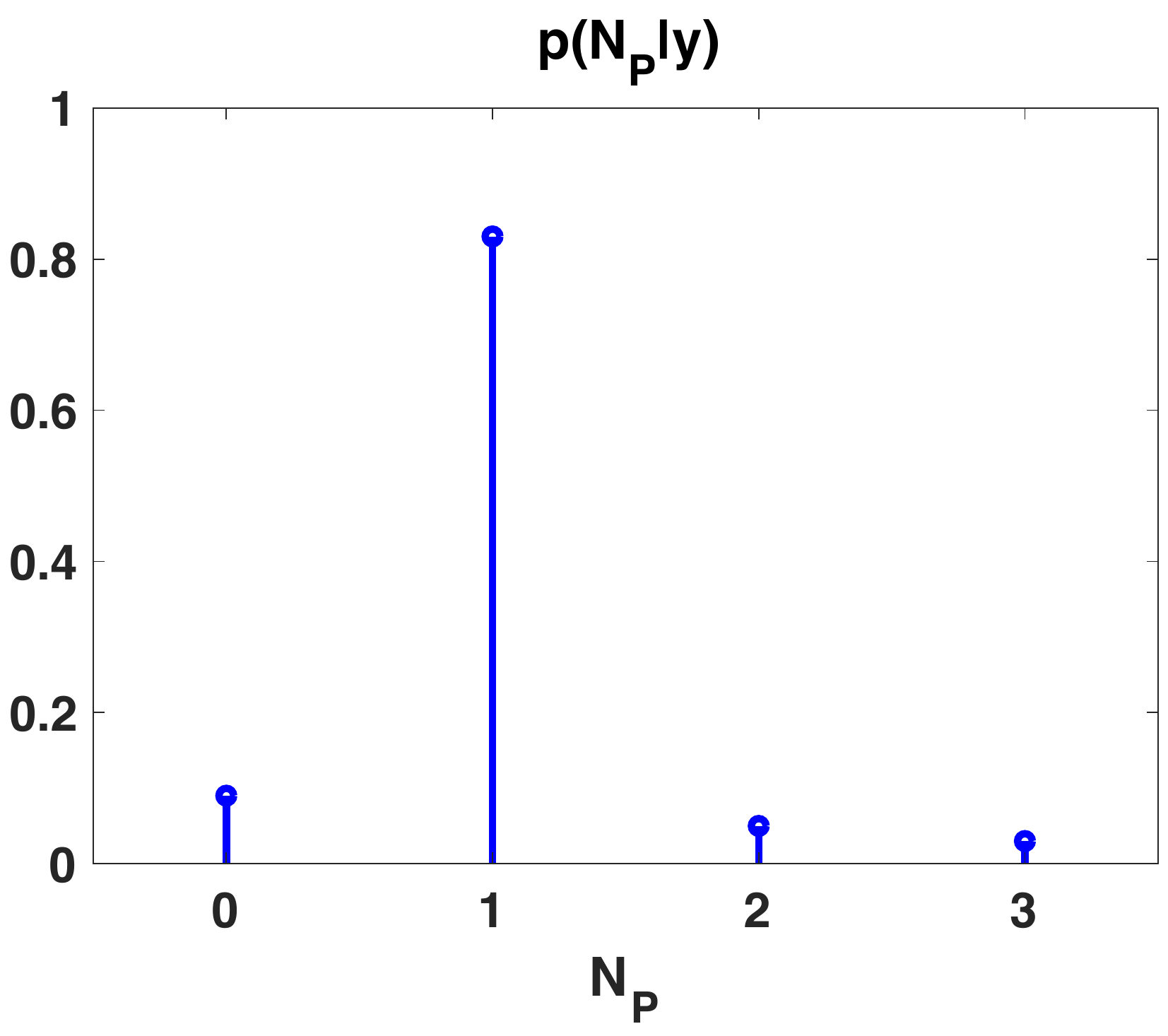}
\includegraphics[width=4cm]{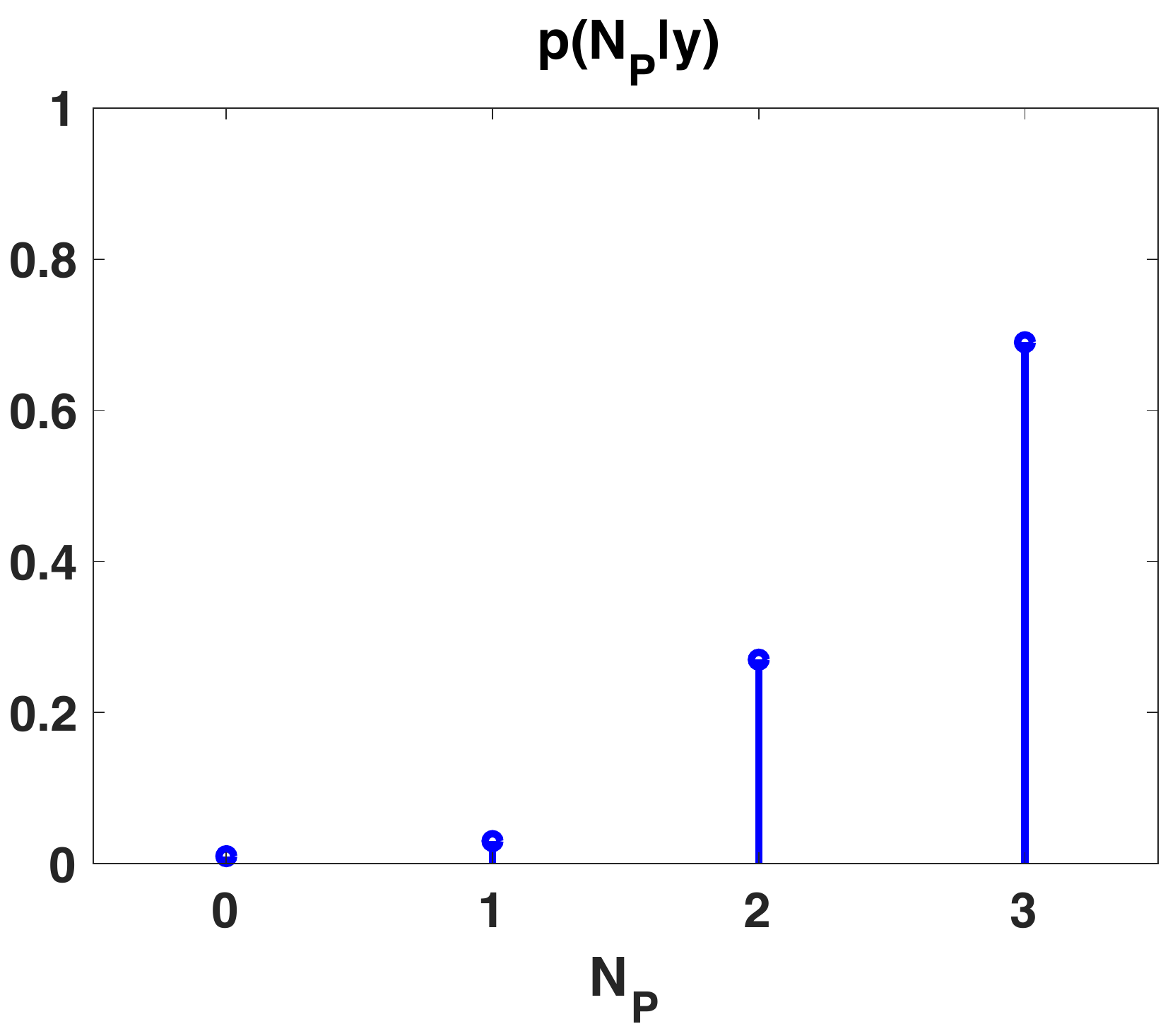}
\caption{{ Approximation of the marginal posterior probability mass of number of planets $N_P$ obtained by using CLAIS with $T=20^5$ and $M=10$ ($N_P=1$ on the left and $N_p=3$ on the right).}}
\label{FigEX_CLAIS}
\end{figure}

\subsection{Compressed Particle Filtering}
\label{CPFsimu}
This section is devoted to { analyzing} the performance of the Compressed Particle Filter (C-PF) described in Table \ref{CPFtable}.
{Given the following} the state-space model 
\begin{equation}
\left\{
\begin{array}{l}
x_{t}=| x_{t-1}|+ v_{t} \\
  y_t=\log(x_{t}^2)+ u_{t}
\end{array}
\right.,
\qquad t=1,\ldots, T,
\end{equation}
where $v_{t}\sim \mathcal{N}(0, 1)$ and $u_{t}\sim \mathcal{N}(0, 1)${, the goal} is to track $x_{t}$ for $T=100$ {time instants}, with a particle filtering algorithm considering $N\in\{100, 1000\}$ particles. We compare the bootstrap particle filter (BPF) \cite{SMC01} and C-PF in terms of the Mean Square Error (MSE) in {the estimation} of $x_{1:T}$. We apply C-PF with different values of $M$ (clearly, with $M\leq N$).  
We consider the deterministic C-MC scheme with a uniform construction P2 of the partition.

Figure \ref{FigEX_CPF} shows the MSE (averaged over $5000$ independent runs) as function of the compression rate, {given by the ratio} $\frac{M}{N}$. The solid lines represent the MSE obtained by the BPF. The dashed line with squares corresponds to the  C-PF (using the deterministic compression) with $N=100$, whereas {the dashed line with circles corresponds} to the  C-PF with $N=1000$. Note that C-PF virtually obtains the same performance of the BPF with approximately $85\%$ {fewer} evaluations of the likelihood function. We recall that the $N$ resampling steps are performing over $M$ particles instead of $N$. Furthermore, fixing the compression {rate of} $\frac{M}{N}$, It is interesting to note that the performance of C-PF improves when $N$ grows. Finally, we have applied an unscented Kalman filter (UKF) \cite{Julier04,Sarkka13bo}{, and} computed its MSE in estimating $x_{1:T}$. C-PF obtains the same or better MSE for $M\geq 20$ when $N=1000$. 

   \begin{figure}[htbp]
\centering
\includegraphics[width=8cm]{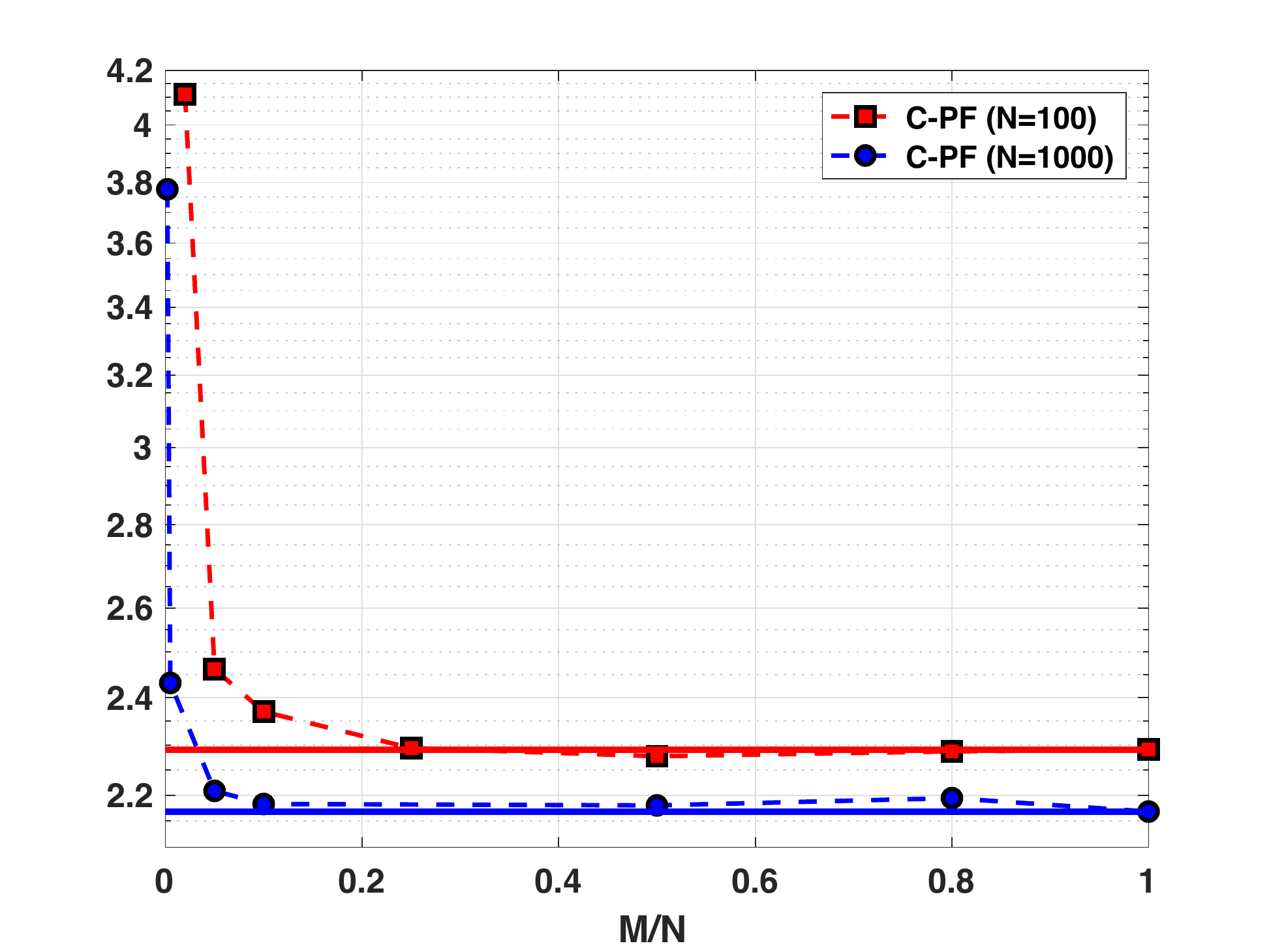}
\caption{{ MSE as function of the compression rate $\frac{M}{N}$. The dashed line with squares corresponds to the  C-PF with $N=100$, whereas  with circles  corresponds to the  C-PF with $N=1000$. The solid lines  corresponds to the bootstrap particle filter with $N=100,1000$.  C-PF virtually obtains  the same performance of the  bootstrap particle filter with approximately $85\%$ less evaluations of the likelihood function.}}
\label{FigEX_CPF}
\end{figure}

\subsection{Tracking with Improved Gaussian particle filtering (I-GPF)}
\label{GPFsect}

In this section, we compare the performance of the benchmark Gaussian particle filter (GPF) with an Improved GPF (I-GPF) method which employs C-MC, described in Table \ref{IGPFtable}. For this {comparison,} we consider {a bearings-only tracking (BOT)} model. The BOT {model arises} in different engineering applications. More specifically, we considers tracking position and velocity of an object moving in a 2D space, ${\bf x}_t=[p_{t,1}, p_{t,2}, v_{t,1},v_{t,2}]^{\top}$ where ${\bf p}_t=[p_{t,1},p_{t,2}]^{\top}$ and ${\bf v}_t=[v_{t,1},v_{t,2}]^{\top}$ are the position and velocity vectors, respectively. The measurements taken by the sensor are the bearings or angles {regarding} the sensor position, contaminated by noise. The range of the object, that is, the distance from the sensor, is not observed. The transition model is
$$
{\bf x}_{t+1}={\bm \Phi} {\bf x}_{t}+{\bm \Gamma}{\bm \eta}_{t+1}, \qquad t=1,...,T,
$$
where
$$
{\bm \Phi}=
\left(
\begin{matrix} 
1 & 0 & 1 & 0 \\
 0 & 1 & 0 & 1 \\
  0 & 0 & 1 & 0 \\
  0 & 0 & 0 & 1\\
\end{matrix}
\right), \quad
{\bm \Gamma}=
\left(
\begin{matrix} 
0.5 & 0  \\
0 & 0.5  \\
1 & 0  \\
0 & 1  \\
\end{matrix}
\right), 
$$
 and ${\bm \eta}_{t+1}=[\eta_{1,t+1},\eta_{2,t+1}]^{\top} \sim \mathcal{N}({\bf 0},\sigma_\eta^2{\bf I})$. The measurements consist of the true bearing of the target contaminated by noise, i.e., the measurement equation is
 $$
 y_t=\arctan\left[\frac{p_{t,1}}{p_{t,2}}\right]+\zeta_{t}
 $$
 where  $\zeta_{i,t}\sim \mathcal{N}(0,\sigma_\zeta^2)$. 
 Note that, with this kind of {observation model}, we obtain no information about the range of the object from the measurement.  At the $t$-th iteration, the GPF algorithm replaces the resampling steps in a standard particle filter by constructing a Gaussian density, given the weighted samples, and sampling from it. In the I-GPF scheme,  the pdf in Eq. \eqref{CMC_PI_KDE_Eq} ($\delta=0.1$) based on the deterministic C-MC, plays the role of the Gaussian density in the standard GPF (see Table \ref{IGPFtable}). We consider a uniform partition P2 (see Section \ref{ChoicePartition}). We generate trajectories of length $T=15$ and measurements from the model with parameters ${\bf x}_0=[-0.05,0.001,0.7,-0.055]$, $\sigma_\eta=0.001$, $\sigma_\eta=0.005$, and number of particles $N=1000$.  We compute the MSE (averaged in the four components) in estimation of ${\bf x}_{1:T}$  (averaged over $10^5$ runs) using GPF and I-GPF with $M\in\{5,10,20\}$. The results  in Table \ref{GPFres}, shown that I-GPF outperforms GPF.
 
  \begin{table}[!h]
\small
{ 
\caption{MSE in estimation of ${\bf x}_{1:T}$ (Ex. in Section \ref{GPFsect}).}
\label{GPFres}
\begin{center}
\begin{tabular}{|c|c|c|c|c|}
   \hline
  Method & $M=5$  & $M=10$ & $M=20$ & $M=30$ \\
   \hline
   \hline
  GPF &  \multicolumn{4}{c|}{ 0.0186}  \\
  \hline
   I-GPF & 0.0157 & 0.0145 & 0.0121 &  0.0098 \\
     \hline  
\end{tabular}
\end{center}
}
\end{table}


\subsection{Application to distributed particle filtering (DPF)}
\label{DistribuidoPF} 
In this section, we consider the nearly coordinated turn model, with state ${\bf x}_t=[p_{t,1},p_{t,2},v_{t,1},v_{t,2}, \gamma_t]^{\top}$, i.e., $d_X=5$,  which contains the position and velocity coordinates (${\bf p}_t=[p_{t,1},p_{t,2}]^{\top}$ and ${\bf v}_t=[v_{t,1},v_{t,2}]^{\top}$), as well as the turn rate $\gamma_t$. Thus,  the transition model is 
$$
{\bf x}_{t+1}=\left(
\begin{matrix} 
1 & 0 & \frac{\sin(\gamma_t)}{\gamma_t}  &  \frac{\cos(\gamma_t)-1}{\gamma_t} & 0 \\
 0 & 1 &  \frac{\cos(\gamma_t)-1}{\gamma_t}    & \frac{\sin(\gamma_t)}{\gamma_t}  & 0 \\
 0 & 0 & \cos(\gamma_t)   & -\sin(\gamma_t)  & 0 \\
 0 & 0 & \sin(\gamma_t)   & \cos(\gamma_t)  & 0 \\
 0 & 0 & 0  & 0  & 1 \\
\end{matrix}
\right) {\bf x}_{t}+{\bm \eta}_{t+1},
$$
where   $t=1,...,T$, ${\bm \eta}_{t+1} \sim \mathcal{N}({\bf 0},{\bf D})$ with ${\bf D}=\mbox{diag}([0.05,0.05,0.04,0.04,0])$, and constant turn rate $w_t=0.139$. 
The measurement equations is 
\begin{eqnarray}
y_{i}=h_i({\bf x}_t)+\zeta_{i,t}
\end{eqnarray}
where $h_i({\bf x}_t)$ represents the specific sensor and $\zeta_{i,t}\sim \mathcal{N}(0,\sigma_i^2)$.
We consider $K$ sensors distributed uniformly in the square region $\mathcal{R}=[-3,3]\times [-3,3]$, with the position denoted as ${\bf r}_i=[r_{i,1},r_{i,2}]^{\top}$, $i=1,...,K$. We consider $4$  different types of sensors: $K/4$ of them are bearing-only sensors,
\begin{eqnarray}
h_i({\bf x}_t)=\arctan\left[\frac{p_{t,1}-r_{i,1}}{p_{t,2}-r_{i,2}}\right],
\end{eqnarray}
with $\sigma_i=0.175$ and  $K/4$ of them are the signal-strength sensors,
\begin{eqnarray}
h_i({\bf x}_t)=\frac{1}{||{\bf p}_t-{\bf r}_i||^2+a},
\end{eqnarray}
with $a=10^{-4}$,  $\sigma_i=2$, $K/4$ of them are the range-measurement sensors,
\begin{eqnarray}
h_i({\bf x}_t)=||{\bf p}_t-{\bf r}_i||,
\end{eqnarray}
with $\sigma_i=0.14$ and $K/4$ of them are the radial-velocity sensors described by
\begin{eqnarray}
h_i({\bf x}_t)=\frac{({\bf p}_t-{\bf r}_i)\cdot {\bf v}_t}{||{\bf p}_t-{\bf r}_i||},
\end{eqnarray}
where  $\sigma_i=0.004$ and $\cdot$ denotes the scalar product.  Each sensor provides one measurement, $y_i$, per iteration. We consider $L\in\{4,8\}$ local processors distributed uniformly in the area $\mathcal{R}$ (in a grid form).
Each sensor transmit to the closest local processor. Hence, each local processor addresses a partial posterior $\bar{\pi}_{\ell}^{(t)}$, using Eqs. \eqref{CMC_PI_KDE_Eq}-\eqref{Sigma_m2} ($\delta=0.1$), with different number of observations. {In the  central node, we perform the information fusion} obtaining $\widehat{\pi}_{\texttt{tot}}^{(t)}$ or, with compression $\widetilde{\pi}_{\texttt{tot}}^{(t)}$. The deterministic C-MC is performed creating a partition of $M$ sets creating a uniform grid strategy P2 suggested in Section \ref{ChoicePartition}. We compare the deterministic C-MC with the ideas proposed in \cite{Oreshkin10} adapted for the central node scenario that coincides with the first method proposed in \cite{EmbaraMCMC} but employed within a particle filtering context.
We set $T=10$,  $K=\{8,16,40\}$ and $M=4$ and compute the MSE in estimation of ${\bf x}_{1:T}$, averaged over $10^4$ independent runs. The results are shown in Table \ref{DPFres}. The proposed technique obtains the smallest MSE since, in general, provides a more robust estimation of ${\bf x}_{1:T}$.

 \begin{table}[!h]
\small
{ 
\caption{MSE in estimation of ${\bf x}_{1:T}$ (Ex. in Section \ref{DistribuidoPF}).}
\label{DPFres}
\begin{center}
\begin{tabular}{|c|c|c|c|}
   \hline
   \multicolumn{2}{|c|}{Method} & $K=8$  & $K=16$ \\
   \hline  
\hline  
\multirow{2}{*}{$L=4$}    &  \cite{Oreshkin10,EmbaraMCMC}  &0.332 &0.186 \\
  & CMC-DPF  &0.161 & 0.095 \\
\hline
\hline
 \multicolumn{2}{|c|}{Method} & $L=4$  & $L=8$ \\
 \hline
\multirow{2}{*}{$K=16$}   &  \cite{Oreshkin10,EmbaraMCMC}  &0.186 & 0.301 \\
  & CMC-DPF  &0.095  & 0.143 \\
  \hline
\end{tabular}
\end{center}
}
\end{table}

}
\vspace{-0.4cm}
\section{Conclusions and future works}
\label{ConSect}

In this work, we have introduced a novel efficient scheme {to} summarize the information provided by Monte Carlo sampling algorithms. This problem is related to the moment matching approach used in different filtering methods but applicable only for certain {target densities.  
 The proposed technique} can be applied in different scenarios, for instance{, in} the distributed inference framework, within advanced particle filtering schemes, or within adaptive Monte Carlo methods. We have introduced three novel Monte Carlo schemes based on C-MC. 
Among them, the C-PF is particularly promising{, since} reducing {considerably} the number of the likelihood evaluations, {C-PF is still able to provide a similar performance of a standard particle filter, with remarkably more evaluations of the likelihood.} In the proposed CLAIS method, we have shown that C-MC can be employed for reducing the computational cost of AIS schemes.
\newline
The C-MC-based algorithms have been tested in six different numerical experiments, considering {several inference problems.} The results have shown that C-MC techniques outperform the corresponding benchmark methods. The deterministic C-MC scheme appears particularly efficient. As future research line, we plan to study the connection between C-MC and sigma-points approaches (see, e.g., in C-PF).  We also plan to analyze the information loss using the Kullback-Leibler (KL) divergence between the C-MC approximation and the true distribution. The  LS-CMC scheme (and its regularized versions) also {deserves further} studies {also} from a theoretical point of view, trying to overcome the difficulty due to the possibility of obtaining negative weights. The joint use of LS-CMC and C-PF will be also investigated. 



\bibliographystyle{plain}
\bibliography{bibliografia}

\appendices

\section{Zero-loss compression for a specific integral $I(h)$}
\label{App1}
{Given a function $h({\bf x})$,} Theorem \ref{Teo1}  states that, with the choice $s_m=\sum_{j\in \mathcal{J}_m} {\bar w}_{m,j} h({\bf x}_j)$ in \eqref{EqSpecH},  we have ${\widetilde I}^{(M)}(f)\equiv {\widehat I}^{(N)}(h)$, {when $f({\bf x})={\bf x}$.}  Indeed, we have 
\begin{eqnarray}
{\widetilde I}^{(M)}(f)&=& \sum_{m=1}^M \widehat{a}_m s_m  \nonumber Ê \\
 &=& \sum_{m=1}^M \widehat{a}_m   \left[\sum_{j\in \mathcal{J}_m} {\bar w}_{m,j} h({\bf x}_j)\right]  \nonumber 
\end{eqnarray}
{and replacing ${\bar w}_{m,j}=\frac{{\bar w}_j}{\widehat{a}_m}$ given in Eq. \eqref{EqREVmagic}, we obtain}
\begin{eqnarray} 
{\widetilde I}^{(M)}(f)&=& \sum_{m=1}^M \widehat{a}_m   \left[\sum_{j\in \mathcal{J}_m} \frac{{\bar w}_j}{\widehat{a}_m} h({\bf x}_j)\right].  \nonumber  \\
&=&  \sum_{m=1}^M \sum_{j\in \mathcal{J}_m} {\bar w}_j h({\bf x}_j)   \nonumber  \\
&=&  \sum_{j=1}^N  {\bar w}_j  h({\bf x}_{j})={\widehat I}^{(N)}(h),  
\end{eqnarray}
{that is the desired result, given in Theorem \ref{Teo1}.}

\vspace{-0.5cm}
\section{Derivation of $c_m(h)$}
\label{App2}

In this Appendix, {the goal is to show} that 
\begin{eqnarray}
c_m(h)&=&\widehat{a}_m^2 \mbox{var}_{\widehat{\pi}_m}[h({\bf s}_m)|\mathcal{S}]  \label{Cmh}\\
         &=&\sum_{i\in\mathcal{J}_m} \bar{w}_i  \sum_{i\in\mathcal{J}_m} \bar{w}_i |h({\bf x}_i)|^2-\Big|\sum_{i\in\mathcal{J}_m} \bar{w}_i h({\bf x}_i)\Big|^2. \nonumber
\end{eqnarray}
First of all, we have
\begin{gather}
\begin{split}
\nonumber
\mbox{var}_{\widehat{\pi}_m}[h({\bf s}_m)|\mathcal{S}]&=\sum_{i\in\mathcal{J}_m} \bar{w}_{m,i} |h({\bf x}_i)|^2-\Big| \sum_{i\in\mathcal{J}_m} \bar{w}_{m,i} h({\bf x}_i)\Big|^2, 
\end{split}
\end{gather}
{ and considering the expressions ${\bar w}_{m,j}=\frac{{\bar w}_j}{\widehat{a}_m}$ given in Eq. \eqref{EqREVmagic} and  $\widehat{a}_m=\sum_{k\in\mathcal{J}_m} \bar{w}_k$ given in Eq. \eqref{EqAmSum}, we obtain} 
\begin{gather}
\begin{split}
\mbox{var}_{\widehat{\pi}_m}[h({\bf s}_m)|\mathcal{S}]&=\frac{ \sum_{i\in\mathcal{J}_m}\bar{w}_{i} |h({\bf x}_i)|^2}{\sum_{k\in\mathcal{J}_m} \bar{w}_k}-\frac{\Big| \sum_{i\in\mathcal{J}_m}\bar{w}_{i} h({\bf x}_i)\Big|^2}{\Big|\sum_{k\in\mathcal{J}_m} \bar{w}_k \Big|^2}.
\end{split}
\end{gather}
Moreover, { again} since $\widehat{a}_m=\sum_{k\in\mathcal{J}_m} \bar{w}_k$ { and replacing above, we can write}
\begin{gather}
\begin{split}
\nonumber
&\widehat{a}_m^2\mbox{var}_{\widehat{\pi}_m}[h({\bf s}_m)|\mathcal{S}]=\\
&=\Big|\sum_{k\in\mathcal{J}_m} \bar{w}_k \Big|^2\frac{ \sum_{i\in\mathcal{J}_m}\bar{w}_{i} |h({\bf x}_i)|^2}{\sum_{k\in\mathcal{J}_m} \bar{w}_k}-\frac{\Big| \sum_{i\in\mathcal{J}_m}\bar{w}_{i} h({\bf x}_i)\Big|^2}{\Big|\sum_{k\in\mathcal{J}_m} \bar{w}_k \Big|^2}, \\
&=\sum_{k\in\mathcal{J}_m} \bar{w}_k \sum_{i\in\mathcal{J}_m}\bar{w}_{i} |h({\bf x}_i)|^2-\Big| \sum_{i\in\mathcal{J}_m}\bar{w}_{i} h({\bf x}_i)\Big|^2,
\end{split}
\end{gather}
that is exactly the expression in Eqs. \eqref{C2} and  \eqref{Cmh}.

\end{document}


%
\title{Supplementary Material of \\ Compressed Monte Carlo \\
with application in particle filtering}
%
%
%

\author{Luca Martino$^{\top}$, V{\'i}ctor Elvira$^*$ \\
{\small $^{\top}$ Dep. of Signal Processing, Universidad Rey Juan Carlos (URJC) and Universidad Carlos III de Madrid (UC3M)} \\
{\small$^*$ IMT Lille Douai, Cit{\'e} Scientifique, Rue Guglielmo Marconi, 20145, Villeneuve dÕAscq 59653, (France)} \\
}
\maketitle

\section{Bias  and variance of stratified estimators}
\label{VarBiasApp}
\vspace{-0.05cm}
{\color{red}Let consider} that $K_m$ samples have been drawn from each sub-region of generic partition of the state space, i.e., $\{{\bf s}_{m,k}\}_{k=1}^{K_m} \sim {\bar \pi}_{m}({\bf x})$, for $m=1,...,M$. Then, the stratified estimator and the corresponding approximation are, respectively, 
\begin{eqnarray*}
&&\widetilde{I}_{V} = \sum_{m=1}^M \bar{a}_m \left[\frac{1}{K_m} \sum_{i=1}^{K_m}    h({\bf s}_{m,i})\right],  \\
 &&{\widetilde \pi}^{(V)}({\bf x}) = \sum_{m=1}^M \sum_{i=1}^{K_m} \frac{\bar{a}_m}{K_m} \delta({\bf x}-{\bf s}_{m,i}),
\end{eqnarray*}
where $\bar{a}_m=\int_{\mathcal{X}_m} {\bar \pi}(x) dx=\frac{Z_m}{Z}${\color{red}, and} $V=\sum_{m=1}^M K_m$ is the total number of generated samples. The stratified estimator is unbiased{\color{red}, i.e.,}  
\begin{eqnarray}
\label{UnbiasedEstPart}
E_{\bar \pi}[\widetilde{I}_{V}]&=&\sum_{m=1}^M \bar{a}_m \left[\int_{\mathcal{X}_m} h({\bf x}) \bar{\pi}_m({\bf x}) d{\bf x}\right], \\
&=&\sum_{m=1}^M \bar{a}_m I_m=I,
\end{eqnarray}
with variance 
\begin{eqnarray}
\mbox{Var}_{\bar \pi}\left[\widetilde{I}_{V}\right]= \sum_{m=1}^M \frac{1}{K_m} \bar{a}_m^2 \sigma_m^2.
\end{eqnarray}
{\color{red} In the expressions above, we have denoted}
$$
I_m=E_{{\bar \pi}_m}\left[h({\bf X})\right]=\int_{\mathcal{X}_m} h({\bf x}) \bar{\pi}_m({\bf x}) d{\bf x},
$$ 
and 
$$
\sigma_m^2= \mbox{var}_{{\bar \pi}_m}\left[h({\bf X})\right] =\int_{\mathcal{X}_m} (h({\bf x})-I_m)^2 \bar{\pi}_m({\bf x}) d{\bf x}.
$$
Namely, $I_m$ and $\sigma_m^2$ represent respectively the mean and the variance of the random variable $h({\bf X})${\color{red}, when} ${\bf X}$ is restricted within {\color{red}the set} $\mathcal{X}_m$ \cite{mcbookOwen,Liu04b,Robert04}. Note that if $K_m=K$ for all $m$, hence $V=MK$, then
\begin{eqnarray}
\mbox{Var}_{\bar \pi}\left[\widetilde{I}_{V}\right]= \frac{1}{K}\sum_{m=1}^M  \bar{a}_m^2 \sigma_m^2.
\end{eqnarray}
 {\color{red}Finally, consider} a proper partition as defined in the main body of this work. {\color{red}In this case,} if $K$ grows the variance of $\widetilde{I}_{V}$ decreases{\color{red}, since} also the area $\bar{a}_m$ corresponding to each becomes smaller and smaller. 

\section{Variance of the random variable $h({\bf X})$} 
Let us recall the definition of the restricted target pdf, ${\bar \pi}_m({\bf x})=\frac{1}{{\bar a}_m} {\bar \pi}({\bf x}) \mathbb{I}_{\mathcal{X}_m}({\bf x})=\frac{1}{Z_m} \pi({\bf x}) \mathbb{I}_{\mathcal{X}_m}({\bf x})$. An interesting expression of the variance of the random variable $h({\bf X})$ can be found, as {\color{red}we show in the following}. Indeed, the variance 
\begin{eqnarray}
\sigma^2=\mbox{var}_{\bar{\pi}}[h({\bf X})]=\int_{\mathcal{D}} (h({\bf x})-I)^2 \bar{\pi}({\bf x}) d{\bf x}, 
\end{eqnarray}
 can expressed as sum of two terms: the first term {\color{red}considers the variance within each the sub-region},
\begin{gather}
\begin{split}
&\sigma_m^2=\mbox{var}_{\bar{\pi}_m}[h({\bf X})]=\int_{\mathcal{X}_m} (h({\bf x})-I_m)^2 \bar{\pi}_m({\bf x}) d{\bf x}, \\
&I_m=E_{\bar{\pi}_m}[h({\bf X})]=\int_{\mathcal{X}_m} h({\bf x}) \bar{\pi}_m({\bf x}) d{\bf x},
\end{split}
\end{gather}
and the second term considers the variance {\color{red}among} the sub-regions, that is $\sum_{m=1}^M  \bar{a}_m (I_m-I )^2$. Namely, we have
\begin{eqnarray}
     \sigma^2 &=&\sum_{m=1}^M  \bar{a}_m \sigma_m^2+\sum_{m=1}^M  \bar{a}_m (I_m-I )^2,
\end{eqnarray}
where we have used the equality $\mbox{var}_{\bar{\pi}}[h({\bf X})]=E_{\bar{\pi}}[\mbox{var}_{\bar{\pi}_m}[h({\bf X})]]+\mbox{var}_{\bar{\pi}}[E_{\bar{\pi}_m}[h({\bf X})]]$  (e.g., see \cite{mcbookOwen, Robert04}). As a consequence, we can also write 
\begin{eqnarray}
\sigma^2\geq \sum_{m=1}^M  \bar{a}_m \sigma_m^2.
\end{eqnarray}
{\color{red}This result} is valid for any kind of partition.

\bibliographystyle{plain}
\bibliography{bibliografia}